\documentclass{aa}

\usepackage[varg]{txfonts}
\usepackage{natbib,graphicx}

\bibpunct{(}{)}{;}{a}{}{,} 
\usepackage{amstext}

\begin{document}

\title{24 synoptic maps of average magnetic field in 296 prominences measured by the Hanle effect during the ascending phase of solar cycle 21\thanks{The 296 prominences solar coordinates and magnetic field vector data are only available in electronic form at the CDS via anonymous ftp to cdsarc.u-strasbg.fr (130.79.128.5) or via http://cdsweb.u-strasbg.fr/cgi-bin/qcat?J/A+A/}}

\author{V. Bommier\inst{1} \and J.L. Leroy\inst{2} \and S. Sahal-Br\'echot\inst{3}}

\institute{LESIA, Observatoire de Paris, Universit\'e PSL, CNRS, Sorbonne Universit\'e, Universit\'e de Paris, 5 place Jules Janssen, F-92190 Meudon, France
\and 65 rue du Ch\^ateau, F-92100 Boulogne-Billancourt, France
\and LERMA, Observatoire de Paris, Universit\'e PSL, CNRS, Sorbonne Universit\'e, 5 place Jules Janssen, F-92190 Meudon, France}

\date{Received ... / Accepted ...}

\abstract
{}
{We present 24 synoptic maps of solar filaments, in which the average unambiguous magnetic field vectors of 296 prominences were determined with Pic-du-Midi observations between 1974 and 1982. This was the ascending phase of cycle 21.}
{The magnetic field was determined by interpreting the Hanle effect, which is observed in the \ion{He}{i} D$_3$ line. Previous results for the prominence field polarity and prominence chirality were applied to solve the fundamental ambiguity. The measurements were averaged in each prominence for accuracy reasons.}
{The result is twofold. First, alternating field directions can be observed from one neutral line to the next. Second, a general field alignment is found along a solar north-south field that is distorted by the differential rotation effect. The numerical data for the prominences and their magnetic field coordinates are provided as online material associated with this paper.}
{}


\offprints{V. Bommier, \email{V.Bommier@obspm.fr}}

\titlerunning{24 synoptic maps of average magnetic fields in 296 prominences}
\authorrunning{V. Bommier et al.}

\maketitle

\section{Introduction}

The first application of the Hanle effect \citep{Hanle-24,Hanle-91} to magnetic field measurements of solar prominences has been made by Hyder in a seminal paper \citep{Hyder-65}, after a theoretical investigation by \citet{Ohman-29}, who showed that the radiation emitted by prominences observed at the limb are linearly polarized by the radiative scattering of the incident solar radiation. As a result, the linear polarization direction would be parallel to the solar limb. This is not the case, however, as observed by \citet{Lyot-34} in the hydrogen H$\alpha$ line of 40 prominences. Further observations of 8 prominences also observed in the \ion{He}{i} D$_3$ line (5875.6 \AA) were reported by \citet{Lyot-36,Lyot-37}. \citet{Hyder-65} summarized all these observations in his Fig. 1, which shows that the linear polarization directions observed by Lyot are rotated with respect to the solar limb. Hyder assigned this rotation of the polarization direction to the Hanle effect, as suggested by \citet{Ohman-29}. The rotation of the scattering linear polarization direction is one of the main features of the Hanle effect. In his Fig. 2, \citet{Hyder-65} also showed that the resulting magnetic field would be consistent with a mainly north-south oriented field that is distorted by the differential rotation effect. The mean magnetic field of the prominence therefore displays a general structure on the Sun. This is the subject of the present paper.

The observations by Bernard Lyot showed that the linear polarization degree of the \ion{He}{i} D$_3$ line is on the order of a few percent. Lyot obtained it with one significant digit. Forty years later, Jean-Louis Leroy undertook new observations of quiescent prominences at the observatory on the Pic-du-Midi with a coronagraph and a polarimeter that were designed and put into operation by \citet{Ratier-75}.  The second significant digit of the \ion{He}{i} D$_3$ polarization degree was determined \citep{Leroy-etal-77}. Leroy planned to observe the magnetic field of the prominence repeatedly in order to detect possible cyclic variations. As a result, 379 prominences were observed between 1974 and 1982. This is the ascending phase of cycle 21.

The first theoretical model of the Hanle effect in solar prominences was developed by \citet{House-70a,House-70b}. However, the atomic density matrix approach developed by \citet{Bommier-SahalB-78} and \citet{Bommier-80} was first able to derive all the required averages on atoms and incident radiation directions from the underlying solar surfaces. This led to the first Hanle effect diagrams of the \ion{He}{i} D$_3$ line of solar prominences \citep{SahalB-etal-77}. These diagrams form a data basis on which linear interpolation was applied to determine the  magnetic fields of the 379 prominences. For accuracy reasons, only 323 prominences were retained at this step. In 6 cases, the observed values were not fit by the diagrams.

However, the obtained magnetic solution is ambiguous. In exact right-angle scattering, two field vectors that are symmetrical with respect to the line of sight cause the same effect on the linear polarization. As a consequence, they remain ambiguous. As discussed in \citet{Bommier-SahalB-78}, this ambiguity is to be related to the very large radiation wavelength with respect to the atomic size. When scattering is not exactly at right angle, the two ambiguous solutions may have slightly different field strengths. However, the ambiguity is always present. The magnetic field vector is not fully determined as long as the ambiguity is not resolved.

Several methods were developed to solve this fundamental ambiguity. One is a comparison of two observations made on consecutive days. The variation in the scattering angle under the effect of solar rotation modifies the symmetry of the solutions. The method is outlined in \citet{Bommier-etal-81} and the results are given in \citet{Bommier-14}. The method was also successfully applied by \citet{Kalewicz-Bommier-19} to a prominence that was observed more recently with spatial resolution with the French TH\'EMIS telescope settled on the European Iza\~na site on the island of Tenerife (Canary Islands, Spain). 

A second method was developed from statistical analysis of the mirror behavior of the ambiguity solution symmetry, together with the previous result of the small angle between the field vector and the long axis of the promincence. This preliminary result was obtained from the Polar Crown prominences of the sample, where the long axis of the prominence, which is the main axis of the associated filament, is aligned with the solar parallel and also with the line of sight. In this case, the angle between the field vector and the long axis of the prominence is the same for the two ambiguous solutions. This angle was found to be on the order of 25$^{\circ}$ \citep{Leroy-etal-83}. The statistical analysis of \citet{Leroy-etal-84} showed that the magnetic field vector of the prominence is mainly of inverse polarity with respect to the neighboring polarities in the photospheric field that are separated by the long axis of the prominence, which coincides with the photospheric neutral line below the filament. 

This result was later confirmed by comparing the ambiguous solutions issued from two spectral lines with a different scattering geometry as a result of the different optical thickness. This technique was applied to 14 prominences observed at the Pic-du-Midi in \ion{He}{i} D$_3$, which is optically thin, and in hydrogen H$\alpha$, which is not optically thin and where the prominence absorption and internal radiation contribution modify the incident radiation anisotropy and scattering. An inverse polarity was found in 12 of the 14 prominences \citep{Bommier-etal-94}. 

Another important point is the horizontality of the magnetic field of the prominences, as shown for the first time by \citet{Athay-etal-83} from the Stokes II spectropolarimeter \citep{Baur-etal-80,Baur-etal-81} observations at Sacramento Peak. The horizontality of the magnetic field of the prominences was later recovered by \citet{Schmieder-etal-14b} from observations with the TH\'EMIS telescope \citep{Schmieder-etal-13,Schmieder-etal-14a}. This result about the field horizontality was applied to the magnetic field determination in the 323 prominences observed at the Pic-du-Midi. The ambiguity was resolved in most cases by selecting the inverse-polarity solution when the two ambiguous solutions correspond to two different polarities. When this method for resolving the ambiguity failed, the ambiguity in a smaller number of cases was resolved by selecting the solution following the chirality law as obtained by \citet{Martin-etal-94}. This law is dextral chirality in the northern hemisphere and sinistral chirality in the southern hemisphere. \citet{Zirker-etal-97} showed that this chirality law is in agreement with the inverse-polarity law for prominences as determined by \citet{Leroy-etal-84}. The chirality law was previously obtained at high latitudes by \citet[][see their Fig. 5]{Leroy-etal-83}.

In nine cases of our sample, the ambiguity could not be solved with the first nor with the second method. In some cases it was not possible to identify the filament associated with the prominence. As a result, we finally determined unambiguous average field vectors in 296 prominences. In the present paper, we publish the synoptic maps of the solar filaments (complemented with the photospheric field polarity and neutral line from the McIntosh maps of NOAA/SEC), on which we have reported the 296 prominence average magnetic field vectors. This confirms the large-scale structure of the prominence and of the solar magnetic field that have first been outlined in Fig. 13 of \citet{Leroy-etal-84}.

In the following, we discuss the method with which we determined the magnetic field  based on the measurements in Sect. \ref{theory} in more detail, and we present the measurement results and synoptic maps in Sect. \ref{results}. We conclude in Sect. \ref{conclusion}.

\section{Hanle effect applied to measurements of the magnetic field in a prominence}
\label{theory}

\subsection{Formalism of the atomic density matrix}

The Hanle effect \citep{Hanle-24,Hanle-91} applies to a spectral line formed by radiative scattering. Except in the case of forward or backward scattering, the line becomes linearly polarized with a polarization direction perpendicular to the scattering plane. From this stage on, the Hanle effect is characterized by two features, namely a depolarization and a rotation of the polarization direction. It is assigned to the partial destruction by the magnetic field of the atomic coherences, or phase relations of different sublevel wavefunctions in the excited state. These phase relations are created by absorption of polarized (or directive) incident radiation \citep{SahalB-81}. However, they are quantitatively described as off-diagonal elements of the atomic density matrix. These off-diagonal elements are called coherences. These coherences are maximum in a zero magnetic field. When the magnetic field separates the sublevels connected by the coherence, it is partially destroyed when the period of the magnetic Larmor precession is shorter than the radiative lifetime. 

When the Larmor frequency is much higher than the inverse radiative lifetime, the coherences are completely destroyed and the linear polarization of the line becomes independent of the magnetic field strength. This is the case of the forbidden lines of the solar corona, where the lifetimes are very long \citep{SahalB-74,SahalB-77}. When $\omega \tau \sim 1$, where $\omega$ is the Larmor pulsation, which is $2 \pi \nu_{\mathrm{L}}$ , where $\nu_{\mathrm{L}}$ is the Larmor frequency, and $\tau$ is the upper level lifetime, more precisely, when $0.1 < \omega \tau < 10$, the depolarization is partial and sensitive to the magnetic field strength. This implies that each spectral line has its own domain in which it is sensitive to the field strength. A table of spectral lines and their sensitivities can be found in \citet[][Table I]{SahalB-81}. This table shows that the \ion{He}{i} D$_3$ line is sensitive to a field strength of about 6 Gauss. The prominence field has often been found to be on this order of magnitude, which makes \ion{He}{i} D$_3$ particularly well suited for quantitative measurements of the magnetic field of a prominence.

In the general case, the line formation description requires solving the statistical equilibrium equations for the atomic levels and sublevels. \citet{SahalB-77} introduced the magnetic sublevels and collisional transitions in the statistical equilibrium equations. As coherences were not accounted for, this formalism was only able to deal with the case of the saturated Hanle effect with $\omega \tau \gg 1$, which is the case of the forbidden coronal lines. The problem was introducing coherences in the statistical equilibrium. In other words, the problem was writing down the statistical equilibrium equations for the full atomic density matrix. This was the PhD work of \citet{Bommier-77}, published in \citet{Bommier-SahalB-78}, which was later extended to higher fields and level crossings by \citet{Bommier-80}. 

As the atomic density matrix contains the average atom properties \citep{SahalB-etal-98}, this formalism for the first time enabled a quantitative calculation of solar radiation scattering in prominences because it was able to integrate partially directive incident radiation, which is fully described by the incident photospheric radiation density matrix \citep{Bommier-SahalB-78}. It was then possible to complete the work initiated by \citet{House-70a,House-70b}. On this occasion, a sign error in the polarization rotation about the magnetic field was corrected. This sign error is present in Fig. 7 of \citet{House-70b} and also in Fig. 2 of \citet{Hyder-65}.

As the experiment by Jean-Louis Leroy used a Lyot filter and not a spectrograph, the line profile was not resolved. As the Hanle effect is an atomic coherence effect that is strongly associated with an atomic level and not with line emission, it is constant throughout the line profile (it is absent from coherent far wings that are due to Rayleigh scattering, however). Therefore line profiles are absent from the above described formalism. After a suggestion by \citet{Landi-83,Landi-84}, the polarized radiative transfer equation was developed for the atomic density matrix. Both system of equations were rederived in the $S$-matrix formalism by \citet[][statistical equilibrium equations]{Bommier-SahalB-91} and \citet[][radiative transfer equation]{Bommier-91}. By repelling the Markov approximation step by step, which has a stopping effect on the perturbation development, the perturbation development of the matter-radiation interaction was pushed forward and finally added up, which led to the introduction of line profiles and partial redistribution in the formalism \citep{Bommier-97a,Bommier-97b,Bommier-16a}.

\subsection{Application to linear polarization in a prominence. Diagrams}

\begin{figure}
\resizebox{\hsize}{!}{\includegraphics{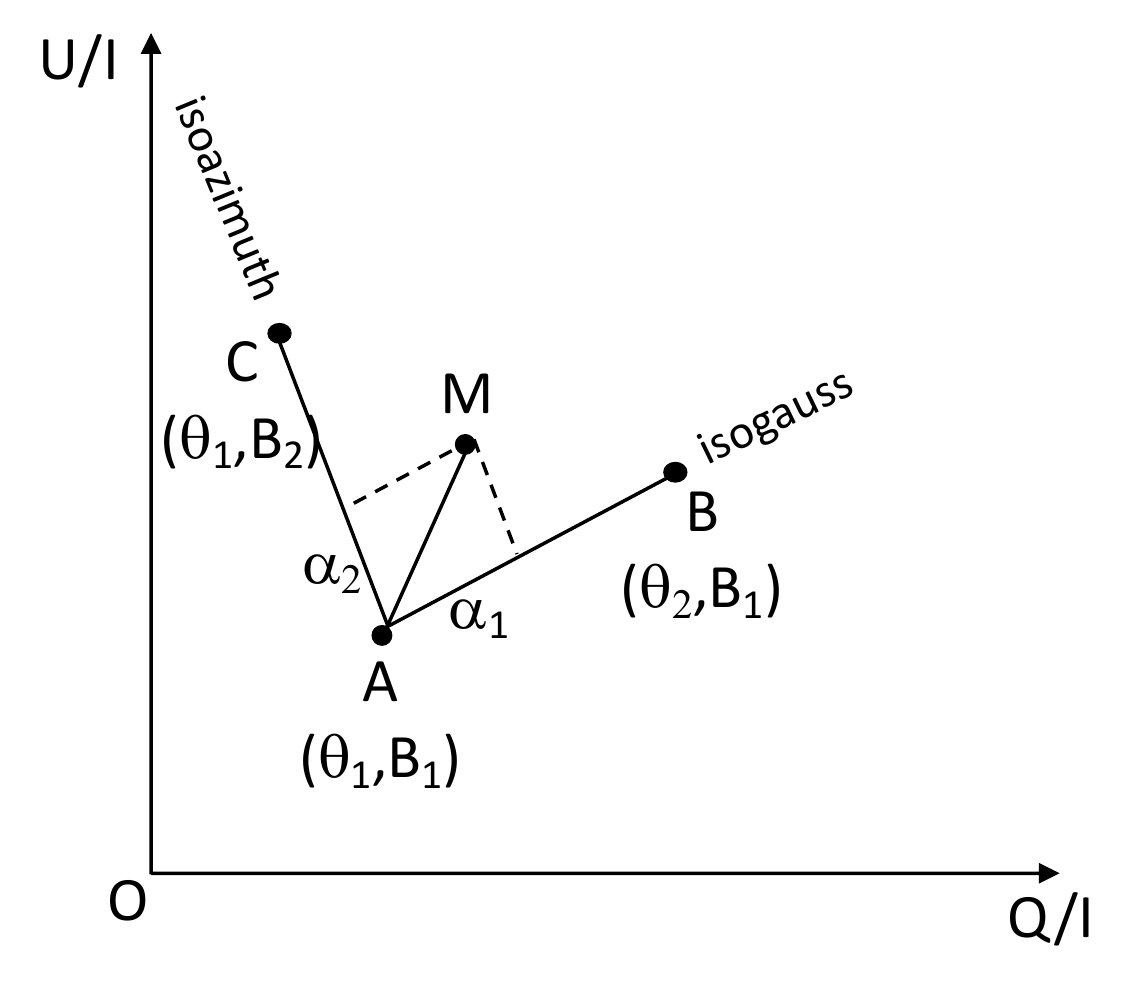}}
\caption{Interpolation method for determining the magnetic field strength $B$ and azimuth $\theta$ of the observed point $M$. This is achieved by determining the affine coordinates $\alpha_1$ and $\alpha_2$ of the $AM$ vector in the affine basis defined by the $AB$ and $AC$ vectors. The points $A, B,\text{and } C$ result from model values. $A$ and $B$ correspond to the same  $B_1$ value (isogauss) of the magnetic
field strength. $A$ and $C$ correspond to the same azimuth $\theta_1$ value (isoazimuth) of the magnetic field.}
\label{interpol}
\end{figure}

As the two main features of the Hanle effect are depolarization and rotation of the linear polarization direction, the Hanle effect is commonly characterized by two parameters, namely the linear polarization degree $p$ and an angle $\varphi$ , which refers to the linear polarization direction with respect to a reference direction $Ox$, which is generally the zero-field polarization direction that results from pure radiative scattering.

These two parameters may be related to two of the Stokes parameters $(I,Q,U,V)$. If $Ox$ is the axis of positive $Q$ and is a reference axis perpendicular to the line of sight $Oz$, $p$ and $\varphi$ may be related to $Q$ and $U$ by
\begin{equation}
\begin{array}{l}
p = \frac{{\sqrt {{Q^2} + {U^2}} }}{I}\\
\cos 2\varphi  = \frac{Q}{{\sqrt {{Q^2} + {U^2}} }}\\
\sin 2\varphi  = \frac{U}{{\sqrt {{Q^2} + {U^2}} }}
\end{array}\ \ .
\end{equation}
It has to be remarked that this equation is able to define $\varphi$ within $\pi$ radians or 180$^\circ$. 

Given these two parameters $p$ and $\varphi$,  the theoretical Hanle effect has commonly been represented by so-called diagrams, which are abaci of the magnetic field effect on linear polarization. Two axes are given in terms of $p$ and $\varphi$ for the vertical and horizontal axis, respectively. A given linear polarization $(p,\varphi)$ is a point in this axis system. Abaci are drawn by plotting the series of points that are obtained by increasing field strengths with fixed field inclination and azimuth, or by an increasing field azimuth for a fixed field strength and inclination. Series of abaci are obtained when the field inclination is varied. In solar prominence, the field inclination is referred to the local solar vertical axis (the solar radius). An abacus but for pure directive scattering can be found in Fig. 7 of \citet{House-70b}. The sign of the polarization rotation is incorrect. The first abaci for the solar prominence case in the \ion{He}{i} D$_3$ line can be found in \citet{SahalB-etal-77}. In Fig. 5 of that paper, observation values are also reported as points. This diagram corresponds to the case of a horizontal magnetic field in a prominence. The observed values fit the theoretical curves very well, in particular in the central empty space, which is the first indication of a horizontal magnetic field in prominences because the non-horizontal diagrams in Fig. 6 of the same paper do not display a central empty space like this. Later on, \citet{Bommier-80} displayed diagrams that were extended to higher fields and to level crossings in the \ion{He}{i} D$_3$ line. Similar diagrams can be found in \citet{Landi-82}. A Hanle diagram is the cover illustration of the monograph "Polarization in Spectral Lines" by \citet{landi-Landolfi-04}.

\subsection{Data inversion by linear interpolation in diagrams}

Observational data can then be interpreted by performing linear interpolation in diagrams as follows. The magnetic field strength $B$ and azimuth $\theta$ of the observed point $M$ are determined by linear interpolation, as schematized in Fig. \ref{interpol}. This is achieved by determining the affine coordinates $\alpha_1$ and $\alpha_2$ of the $AM$ vector in the affine basis defined by the $AB$ and $AC$ vectors, in such a way that $AM = \alpha_1 AB + \alpha_2 AC$ (in vectors). The points $A, B,\text{and } C$ result from model values. $A$ and $B$ are obtained from the same value of the magnetic field strength $B_1$  and from two different field azimuth values $\theta_1$ and $\theta_2$  in such a way that they belong to the same isogauss curve of the diagram. $A$ and $C$ are obtained from the same value of the magnetic field azimuth $\theta_1$  and from two different field strength values $B_1$ and $B_2$  in such a way that they belong to the same isoazimuth curve of the diagram. $A, B, \text{and }C$ are the model points closest to M, in such a way that both $\alpha_1 < 1$ and $\alpha_2 < 1$. Then, the magnetic field strength in $M$ is evaluated as $B = B_1 + \alpha_2 (B_2 - B_1)$ and the magnetic field azimuth is evaluated as $\theta = \theta_1 + \alpha_1 (\theta_2 - \theta_1)$. The uncertainty on the magnetic field strength and azimuth determination may accordingly be derived from the polarimetric accuracy. The diagram may be plotted in Stokes $Q/I$ and $U/I$ coordinates as well, as represented in Fig. \ref{interpol}, as we did in fact in practice.

\subsection{Horizontality of the magnetic field in a prominence}

However, as two linear polarization parameters are measured when three coordinates of the magnetic field vector are searched for, the field solution is not unique and can be represented as a function of the inclination angle $\psi$ between the local solar radius and the magnetic field vector \citep[][Fig. 3]{Bommier-etal-81}. A diagram is plotted, and the above interpolation is performed for each $\psi$ value.

The full determination of the field vector can be achieved by crossing solutions issued from two spectral lines of different sensitivity to the Hanle effect. The \ion{He}{i} D$_3$ line itself is comprised of two fine-structure components, a major one and a minor one. The major component is comprised of five unresolved fine-structure components $3d^{3}D_{3,2,1}\rightarrow2p^{3}P_{2,1}$ , whereas the minor component is comprised of the single $3d^{3}D_{1}\rightarrow 2p^{3}P_{0}$ component. These two components have an unequal sensitivity to the Hanle magnetic field, as shown in Table I of \citet{SahalB-81}. This results from a different lifetime and from different Land\'e factors. As a consequence, they could provide a full magnetic field vector solution. However, they are very close and partially overlap. Their respective observation then requires a spectrograph such as the Stokes II instrument was \citep{Baur-etal-80,Baur-etal-81}. Notwithstanding the problem of separating these two components for a polarization analysis, \citet{Athay-etal-83} and \citet{Querfeld-etal-85} were able to determine the full field vector in 13 and 2 prominences, respectively. They showed that the magnetic field in promincences is mainly horizontal. 

The Pic-du-Midi data reported in this paper, in contrast, were observed through a Lyot filter. The two components were then not resolved and only two parameters were measured, which are the linear polarization degree and direction of the full line. The above result of a horizontal magnetic field in prominences was then applied for the analysis. At the end of the observation campaign, the polarimeter was modified to quasi-simultaneously observe the hydrogen H$\alpha$ and H$\beta$ lines. By using both \ion{He}{i} D$_3$ and hydrogen H$\beta$, \citet{Bommier-etal-86a} were able to again find nearly horizontal magnetic field vectors in 14 prominences.

\citet{Schmieder-etal-14b} confirmed the horizontality for the magnetic field in prominences from a resolved \ion{He}{i} D$_3$ observation interpreted with a principal component analysis (PCA) applied to the polarization profiles \citep{LopezA-Casini-02,LopezA-Casini-03,Casini-etal-03,Casini-etal-05,Casini-etal-09}. Further investigations were possible inside the fine structure of the prominences with a better spatial resolution \citep{Levens-etal-16a,Levens-etal-16b,Levens-etal-17,Schmieder-etal-17}.

The \ion{He}{i} D$_3$ line is absent from the incident photospheric spectrum. As a consequence, there is neither Doppler-dimming nor Doppler-brightening effect in this line, and the prominence velocity field may be ignored in the analysis, which simplifies the problem and yet increases the interest in this line.

\section{Measurement results for our sample}
\label{results}

\subsection{Data description}

From our full sample of 3297 measurements achieved in 379 quiescent prominences observed in \ion{He}{i} D$_3$ at the Pic-du-Midi during the ascending phase of Cycle 21 (1974-1982), we discarded the prominences for which the identification of the neutral line was doubtful (64 prominences). Several scattered and surprisingly rather uniform recordings were made in each prominence, as represented in Fig. 1 of \citet{Leroy-etal-77}, through a 5 arcsec wide pinhole. However, the different measurements in each prominence were finally averaged for accuracy reasons. We discarded the measurements that were not precise enough (inaccuracy higher than $5 \times 10^{-3}$ for the average linear polarization degree, or 10$^{\circ}$ for the average linear polarization direction). The result is a unique determination of the average magnetic field vector of 296 quiescent prominences, corresponding to 2390 measurements.  As explained above, the inversion was obtained by linear interpolation in the horizontal field diagram \citep[Fig. 5 of][later extended to higher field strengths]{SahalB-etal-77}. In six cases only did the observation fall outside of the diagram. The numerical values of each prominence location on the solar surface and the observed magnetic field vector coordinates are provided in a table as online material associated with this paper and available at CDS (see paper title footnote). Recent observations with a better spatial resolution \citep{Schmieder-etal-14b,Levens-etal-16a,Levens-etal-16b,Kalewicz-Bommier-19} display a rather homogeneous magnetic field across quiescent prominences, so that the consideration of their average field makes sense.

\subsection{Synoptic maps}

The object of the present paper is to plot these average magnetic field vectors of 296 prominences in synoptic maps of the solar filaments. A prominence is a filament observed at the limb. A filament is a stretched structure along a neutral line of the photospheric magnetic field. This neutral line separates the positive from the negative photospheric polarity. We used the 1974-1982 synoptic maps prepared at the Meudon Observatory by M.J. Martres and I. Soru-Escaut, and we complemented them with the magnetic information, neutral lines, and photospheric polarities from the McIntosh maps (NOAA/SEC). These synoptic maps are provided in Figs. \ref{1616}-\ref{1722}. Each synoptic map displays the whole solar surface observed close to the disk center and the central meridian throughout one full solar rotation, which is 27.28 days long on average. The Carrington longitude is the horizontal axis and the solar latitude is the vertical axis of the plot. The neutral lines of the photospheric magnetic field taken from the McIntosh maps are added as green lines, and the photospheric magnetic field polarities also taken from the McIntosh maps are referred to with plus (for positive polarity) and minus (for negative polarity) signs on either side of the neutral lines.

\subsection{Resolving the ambiguity of the magnetic field vector}

\begin{figure}
\resizebox{\hsize}{!}{\includegraphics{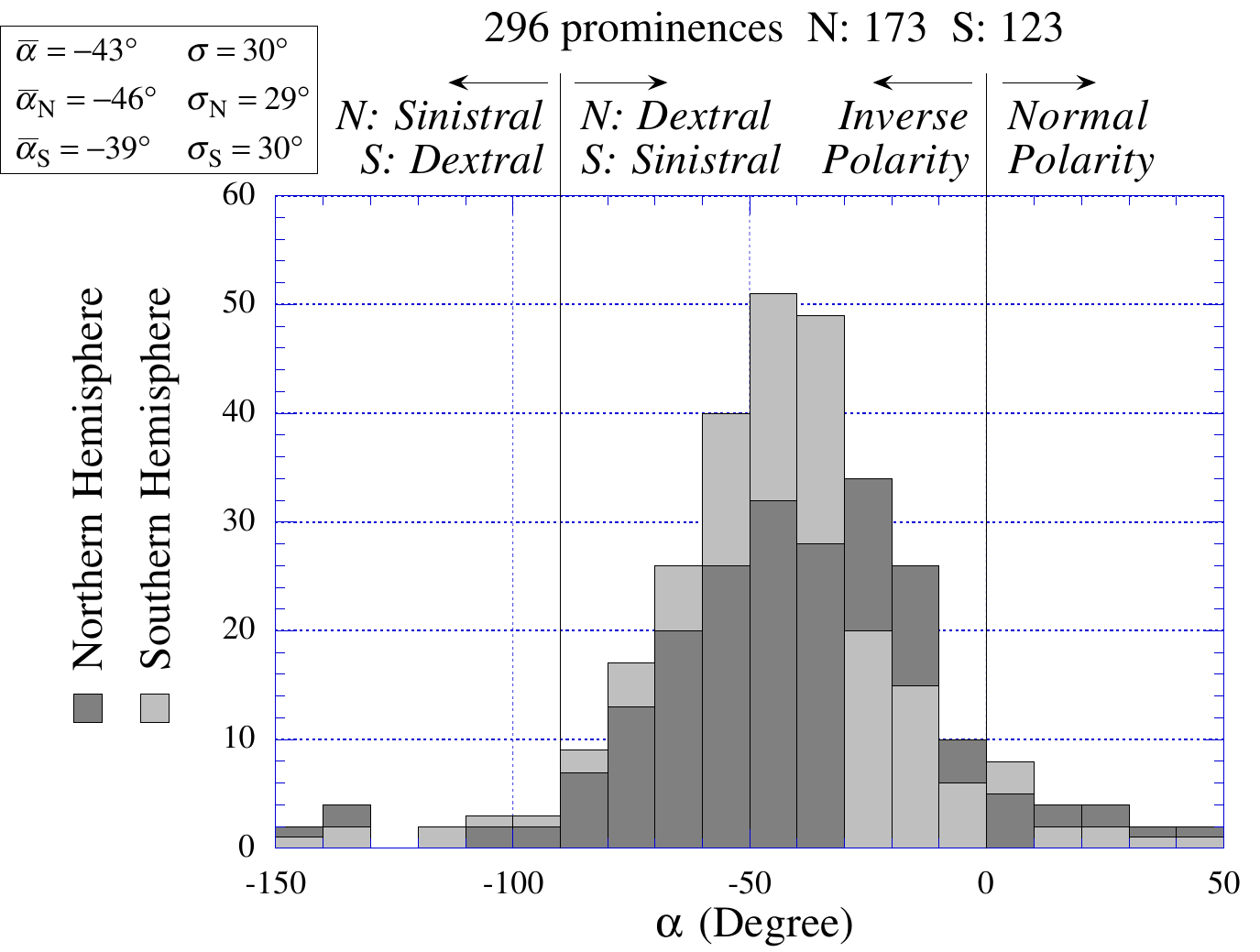}}
\caption{Histogram of the $\alpha$ angle between the magnetic field vector and the long axis of the filament. The angle is oriented as in Fig. 4 of \citet{Leroy-etal-84}. Positive $\alpha$ angles correspond to normal polarity. Negative $\alpha$ angles correspond to inverse polarity. In the northern hemisphere, the filament is dextral when $\alpha$ lies inside the interval $\left[ -90{{}^{\circ }},+90{{}^{\circ }}\right]$, and sinistral outside (and the reverse for the southern hemisphere). The overlined values in the top left corner are the average values, global and for each hemisphere. The associated standard deviations are also provided.}
\label{histogram}
\end{figure}

However, the magnetic field solution is ambiguous, as already discussed. In a first step, the ambiguity was removed in all cases in which the two ambiguous solutions have opposite polarities by selecting the inverse polarity solution following \citet{Leroy-etal-84}. We call this method the polarity method. The polarity of the prominence magnetic field is determined with respect to the neighboring photospheric field polarities. The polarity method enabled resolving the ambiguity for 264 of the full sample of 296 prominences. As the two ambiguous solutions are nearly symmetrical with respect to the line of sight, the method fails when the neutral line lies along a meridian because in this case, the two ambiguous solutions have the same polarity. As observed by \citet{Zirker-etal-97}, most of these 264 prominences to which the polarity method was successfully applied are found to also obey the chirality law of \citet{Martin-etal-94}. This law is dextral chirality in the northern hemisphere and sinistral chirality in the southern hemisphere. Therefore, in a second step the ambiguity was removed for the 32 remaining prominences to which the polarity method does not apply, by applying the chirality law instead. The chirality law was previously obtained at high latitudes by \citet[][see their Fig. 5]{Leroy-etal-83}.

As the two ambiguous solutions are nearly symmetrical with respect to the line of sight, the chirality method fails when the neutral line lies along a parallel, and the polarity method fails when the neutral line lies along a meridian. Where one of the two methods fails, the other method applies. Thus, the two methods are complementary, and when both methods apply, their results agree for nearly all the cases. However, it must be emphasized that the chirality method is derived from the results of the polarity method. Moreover, by using this procedure, a few normal polarity prominences were derived with the chirality method, and accordingly, a few prominences are found not to obey the chirality law, by using the polarity method to solve the ambiguity. Such exceptions are preferentially found in places where the general direction of a neutral line changes.

Figure \ref{histogram} shows the histogram of the angle $\alpha$ between the field vector and the neutral line, oriented as defined by \citet[][see Fig. 4]{Leroy-etal-84}. Positive and negative $\alpha$ angles correspond to normal and inverse polarity, respectively. In Fig. \ref{histogram} the two hemispheres are distinguished. In the northern hemisphere, the filament is dextral when $\alpha$ lies inside the interval $\left[ -90{{}^{\circ }},+90{{}^{\circ }}\right]$, and sinistral outside (the reverse for the southern hemisphere). The average value $\overline{\alpha } =-43{{}^{\circ }}$ and standard deviation $\sigma =30{{}^{\circ }}$ do not
contradict previous results $\overline{\alpha }=-25{{}^{\circ }}$ and $\sigma =38{{}^{\circ }}$ \citep{Leroy-etal-84}, due to the measurement (and other causes of) inaccuracies.

The prominences we observed were found to be of totally inverse polarity. We mean that an inverse polarity is found for the axial and transverse components of their magnetic field vector.

\subsection{Discussion of the results}

\begin{figure}
\resizebox{\hsize}{!}{\includegraphics{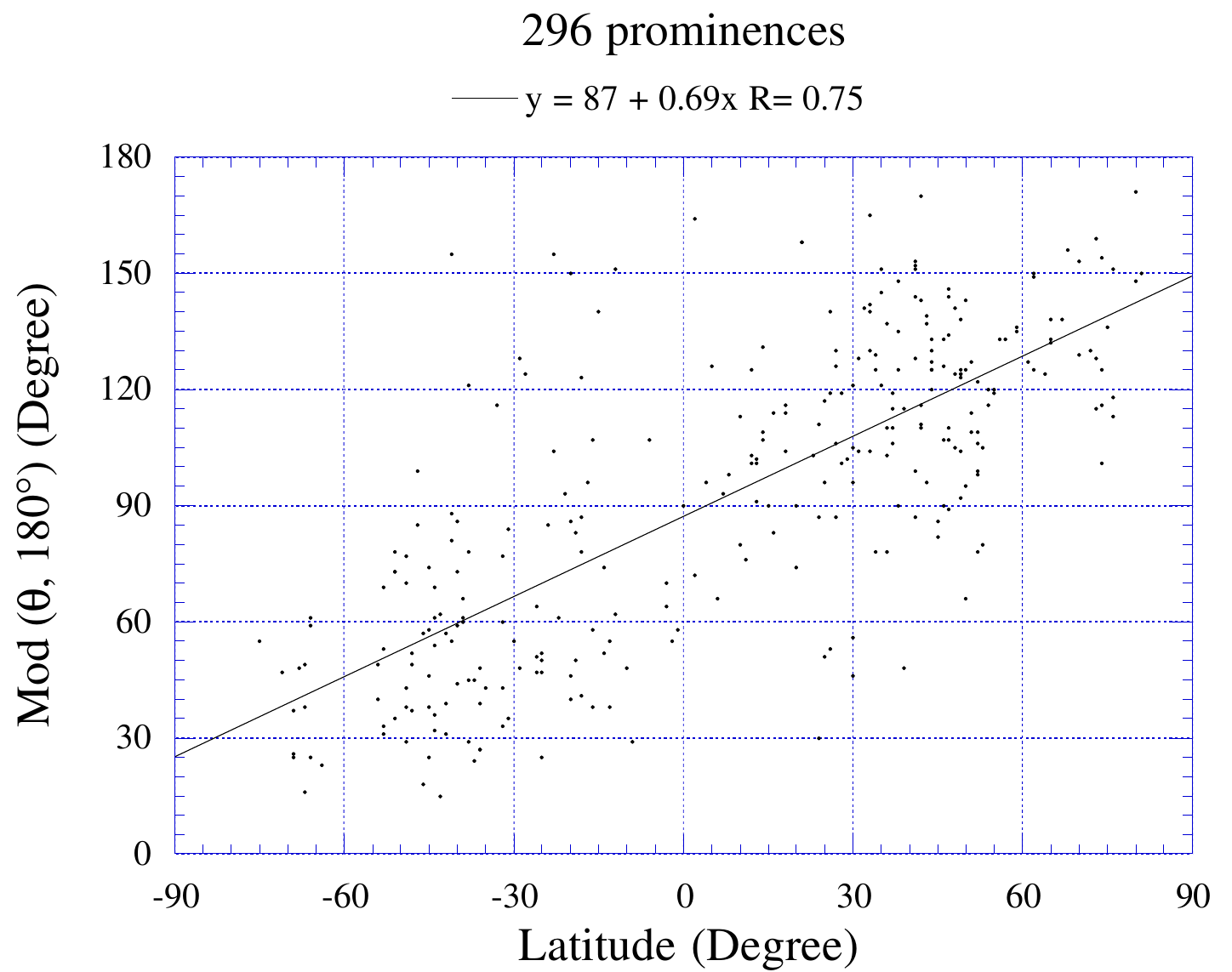}}
\caption{Angle $\theta$ (modulo $180{^{\circ}}$) between the average field vector of each prominence and the W-oriented  solar parallel as a function of the solar latitude.  Top: Equation and 
correlation coefficient of the linear least-squares fit.}
\label{azimuths}
\end{figure}

\begin{figure*}
\resizebox{\hsize}{!}{\includegraphics{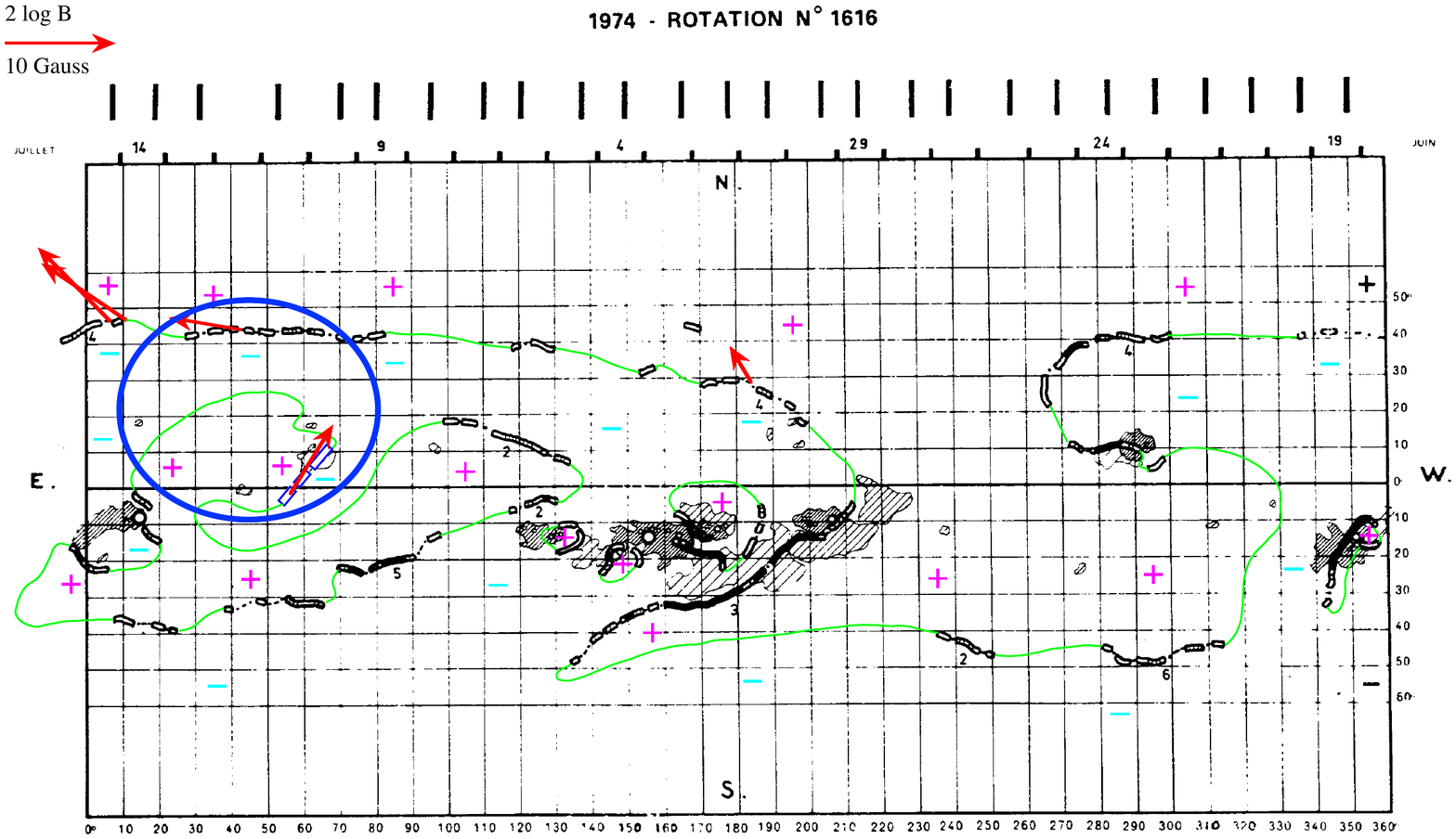}}
\caption{Synoptic map of rotation 1616. Horizontal axis: Carrington longitude. Vertical axis: Solar latitude. The neutral lines of the photospheric magnetic field taken from the McIntosh maps are added as green lines, and the polarities of the photospheric magnetic field also taken from the McIntosh maps are referred to with plus (for positive polarity, in purple) and minus (for negative polarity, in cyan) signs on either side of the neutral lines. Each measured magnetic field vector (average on an observed prominence) is represented by a vector of length $2\mathrm{log}B$. This length for $B=10$ Gauss is plotted in the upper left corner. The magnetic field arrow is plotted in red when the ambiguity is resolved by applying the polarity method. The magnetic field arrow is plotted in orange when the ambiguity is resolved by applying the chirality method. The blue circle embraces parallel filament lines with alternating long-axis magnetic field components, as in Fig. 5 of \citet{Leroy-etal-83}.}
\label{1616}
\end{figure*}

\begin{figure*}
\resizebox{\hsize}{!}{\includegraphics{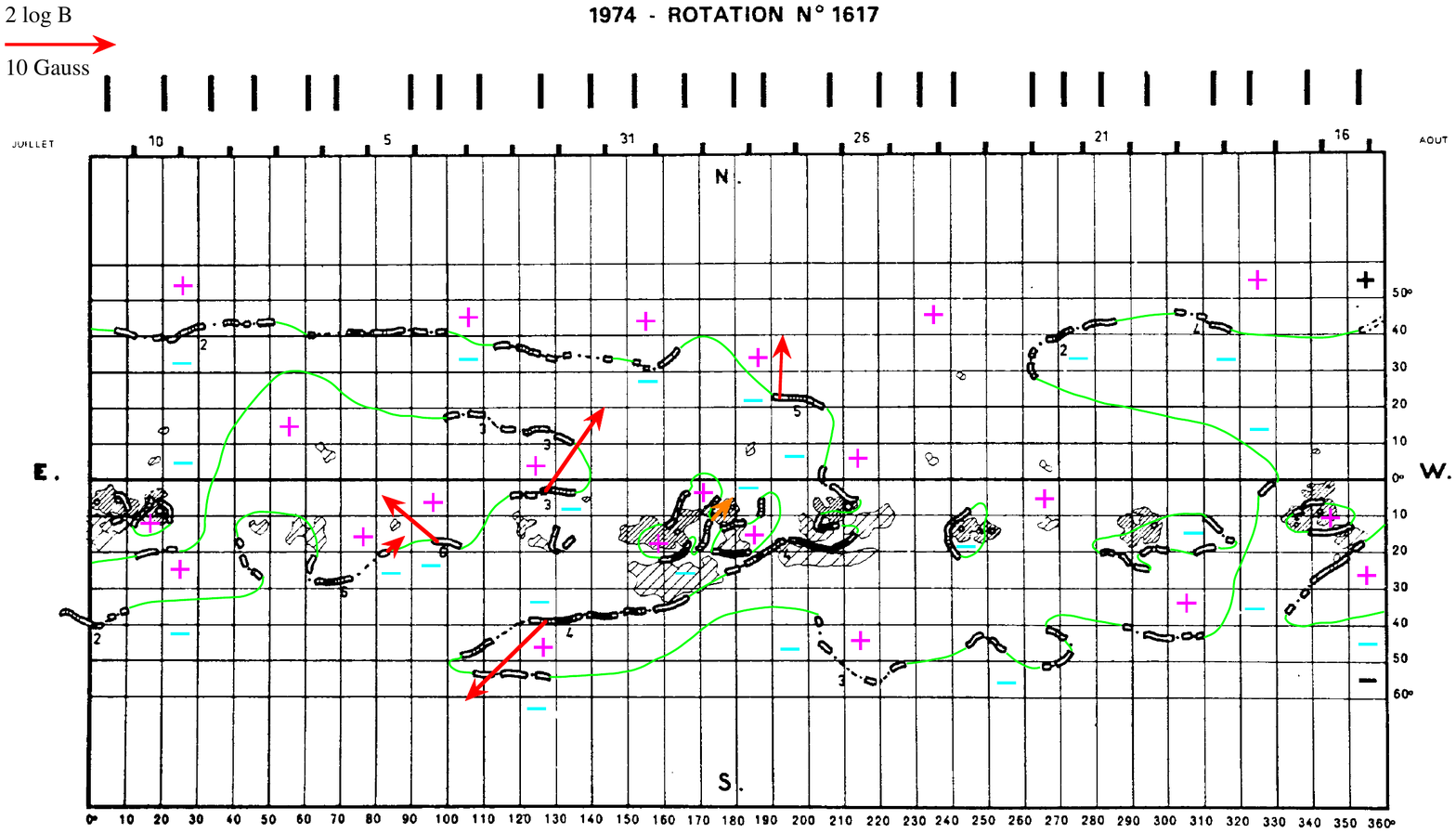}}
\caption{Synoptic map of rotation 1617. Horizontal axis: Carrington longitude. Vertical axis: Solar latitude. The neutral lines of the photospheric magnetic field taken from the McIntosh maps are added as green lines, and the polarities of the photospheric magnetic field also taken from the McIntosh maps are referred to with plus (for positive polarity, in purple) and minus (for negative polarity, in cyan) signs on either side of the neutral lines. Each measured magnetic field vector (average on an observed prominence) is represented by a vector of length $2\mathrm{log}B$. This length for $B=10$ Gauss is plotted in the upper left corner. The magnetic field arrow is plotted in red when the ambiguity is resolved by applying the polarity method. The magnetic field arrow is plotted in orange when the ambiguity is resolved by applying the chirality method..}
\label{1617}
\end{figure*}

\begin{figure*}
\resizebox{\hsize}{!}{\includegraphics{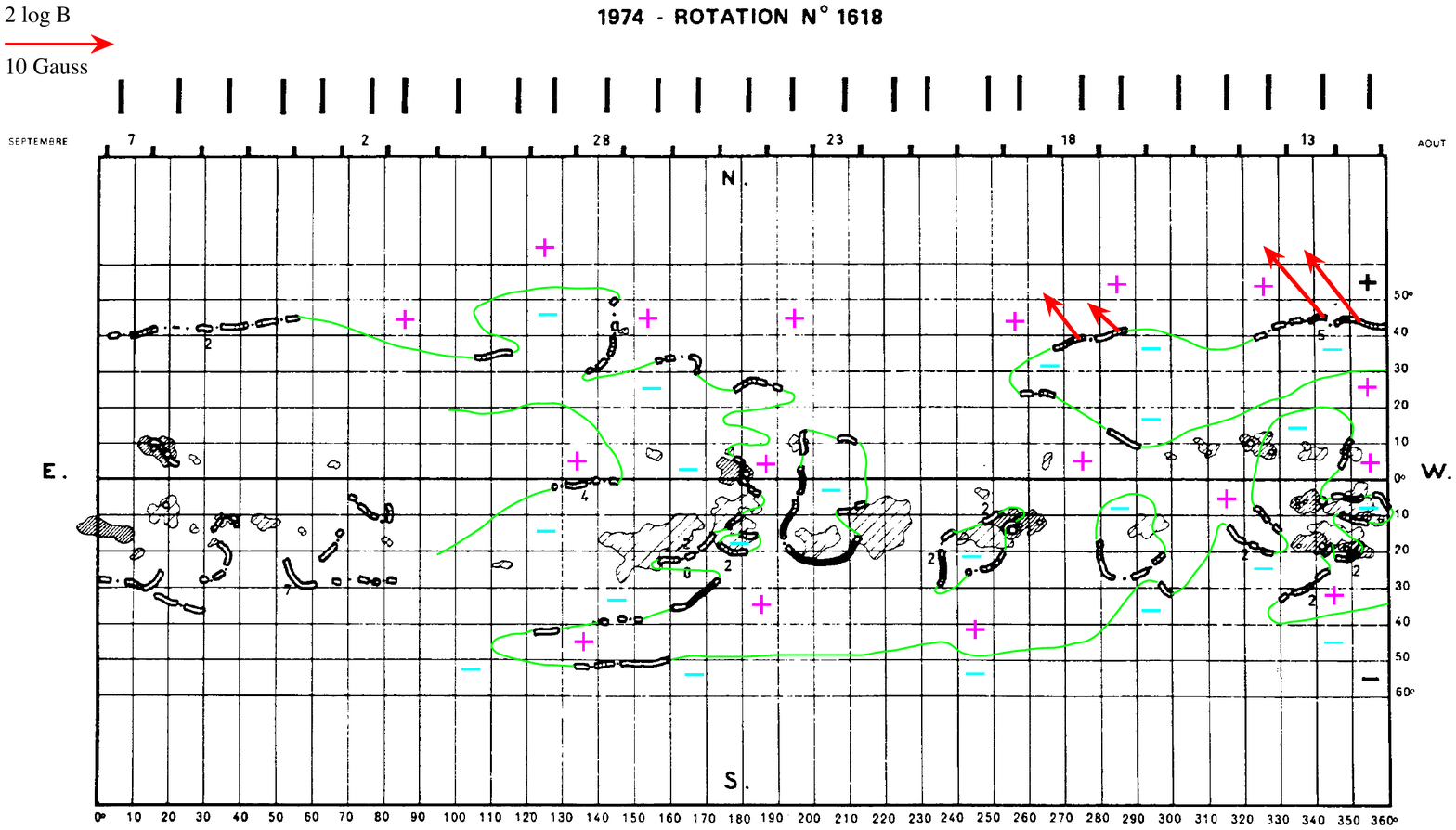}}
\caption{Synoptic map of rotation 1618. Horizontal axis: Carrington longitude. Vertical axis: Solar latitude. The neutral lines of the photospheric magnetic field taken from the McIntosh maps are added as green lines, and the polarities of the photospheric magnetic field also taken from the McIntosh maps are referred to with plus (for positive polarity, in purple) and minus (for negative polarity, in cyan) signs on either side of the neutral lines. Each measured magnetic field vector (average on an observed prominence) is represented by a vector of length $2\mathrm{log}B$. This length for $B=10$ Gauss is plotted in the upper left corner. The magnetic field arrow is plotted in red when the ambiguity is resolved by applying the polarity method. The magnetic field arrow is plotted in orange when the ambiguity is resolved by applying the chirality method.}
\label{1618}
\end{figure*}

\begin{figure*}
\resizebox{\hsize}{!}{\includegraphics{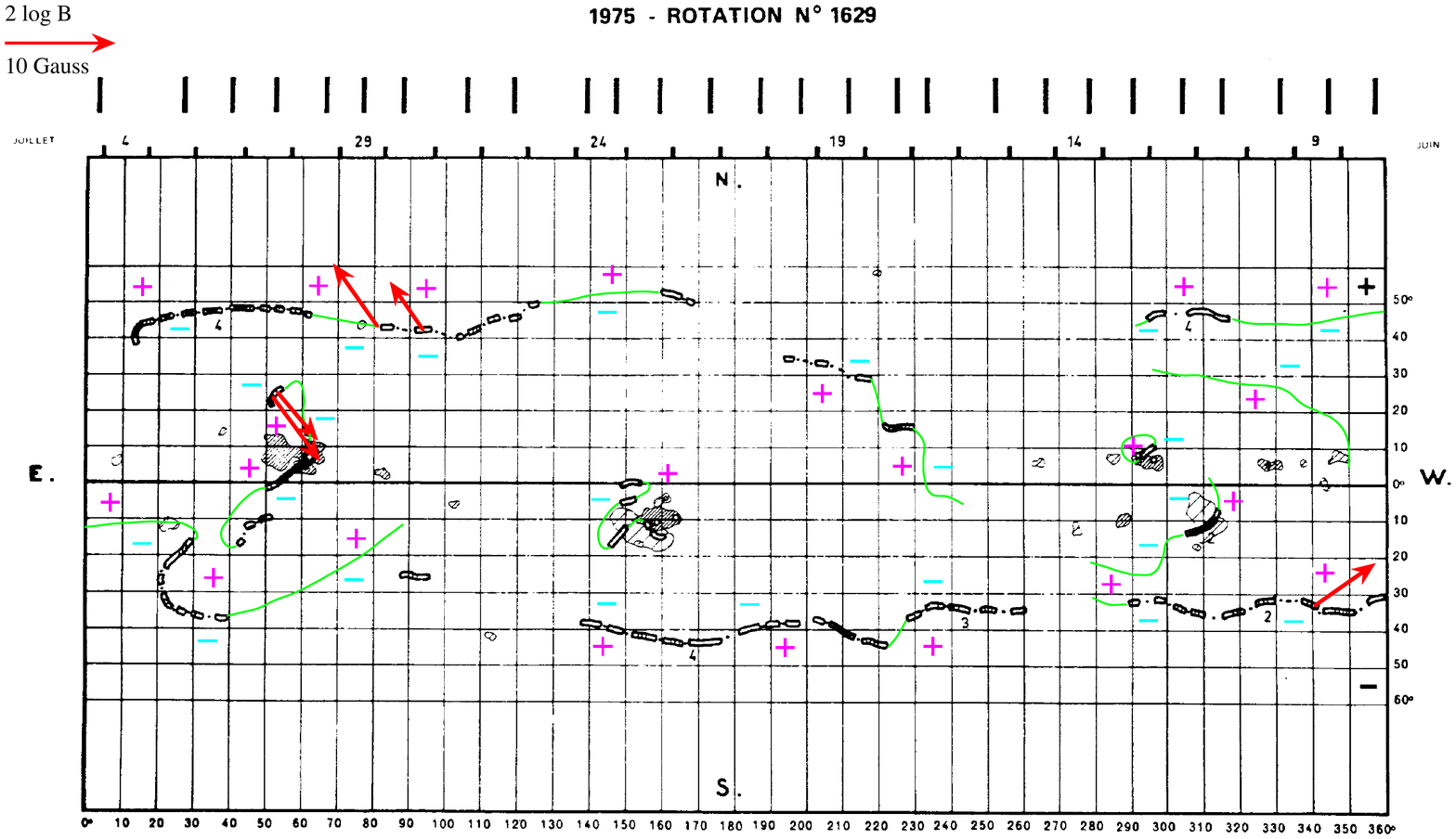}}
\caption{Synoptic map of rotation 1629. Horizontal axis: Carrington longitude. Vertical axis: Solar latitude. The neutral lines of the photospheric magnetic field taken from the McIntosh maps are added as green lines, and the polarities of the photospheric magnetic field also taken from the McIntosh maps are referred to with plus (for positive polarity, in purple) and minus (for negative polarity, in cyan) signs on either side of the neutral lines. Each measured magnetic field vector (average on an observed prominence) is represented by a vector of length $2\mathrm{log}B$. This length for $B=10$ Gauss is plotted in the upper left corner. The magnetic field arrow is plotted in red when the ambiguity is resolved by applying the polarity method. The magnetic field arrow is plotted in orange when the ambiguity is resolved by applying the chirality method.}
\label{1629}
\end{figure*}

\begin{figure*}
\resizebox{\hsize}{!}{\includegraphics{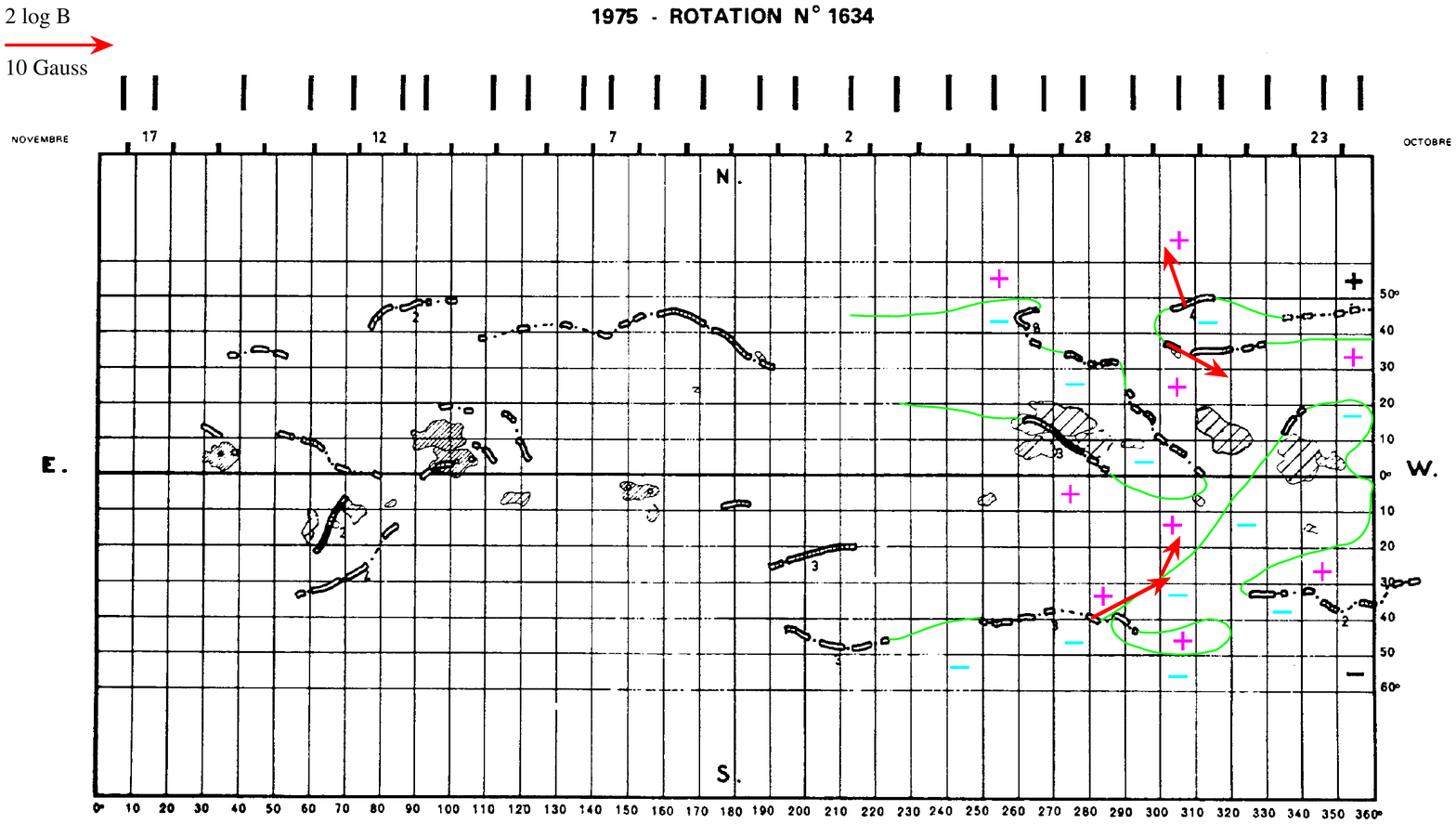}}
\caption{Synoptic map of rotation 1634. Horizontal axis: Carrington longitude. Vertical axis: Solar latitude. The neutral lines of the photospheric magnetic field taken from the McIntosh maps are added as green lines, and the polarities of the photospheric magnetic field also taken from the McIntosh maps are referred to with plus (for positive polarity, in purple) and minus (for negative polarity, in cyan) signs on either side of the neutral lines. Each measured magnetic field vector (average on an observed prominence) is represented by a vector of length $2\mathrm{log}B$. This length for $B=10$ Gauss is plotted in the upper left corner. The magnetic field arrow is plotted in red when the ambiguity is resolved by applying the polarity method. The magnetic field arrow is plotted in orange when the ambiguity is resolved by applying the chirality method.}
\label{1634}
\end{figure*}

\begin{figure*}
\resizebox{\hsize}{!}{\includegraphics{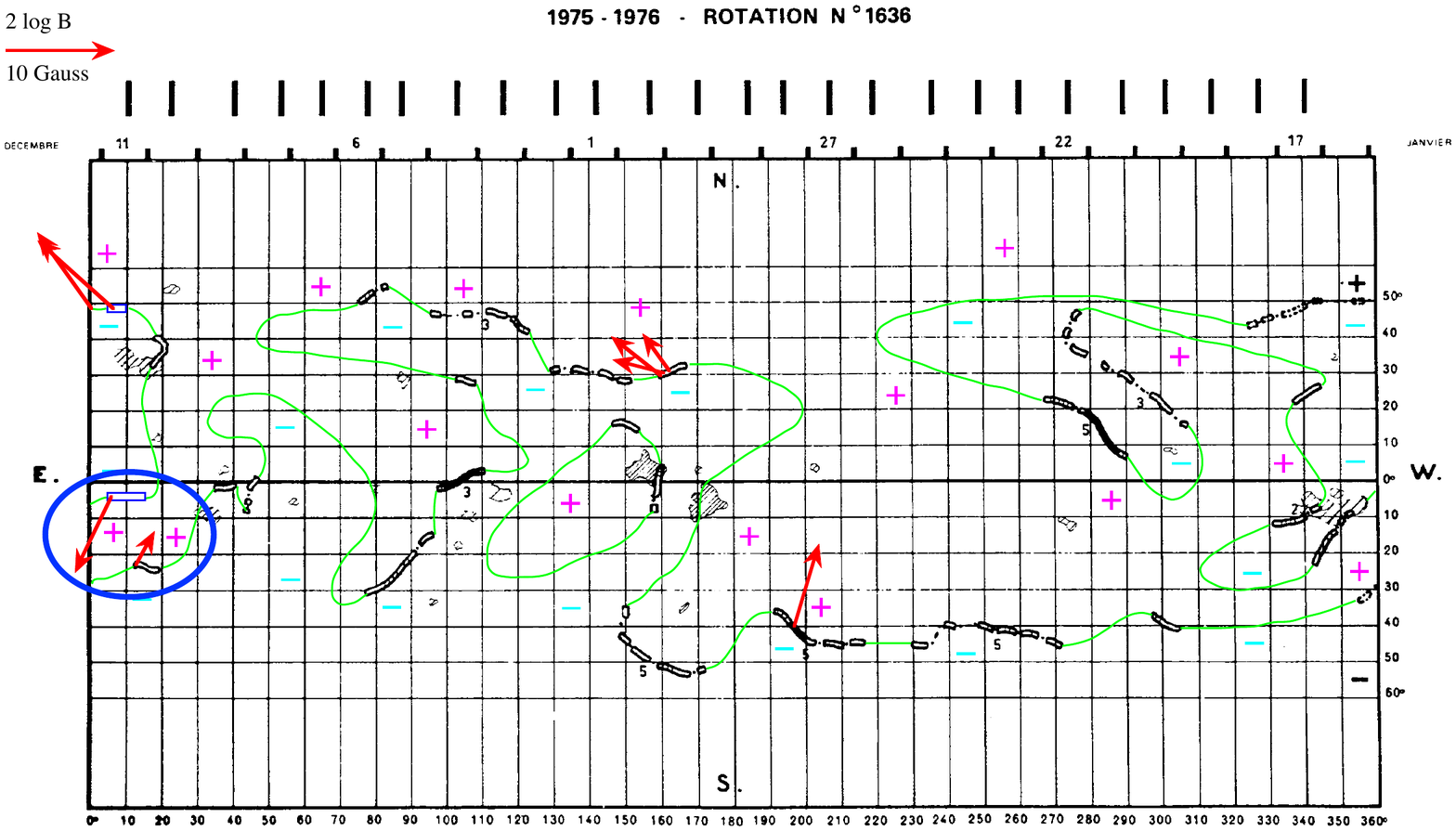}}
\caption{Synoptic map of rotation 1636. Horizontal axis: Carrington longitude. Vertical axis: Solar latitude. The neutral lines of the photospheric magnetic field taken from the McIntosh maps are added as green lines, and the polarities of the photospheric magnetic field also taken from the McIntosh maps are referred to with plus (for positive polarity, in purple) and minus (for negative polarity, in cyan) signs on either side of the neutral lines. Each measured magnetic field vector (average on an observed prominence) is represented by a vector of length $2\mathrm{log}B$. This length for $B=10$ Gauss is plotted in the upper left corner. The magnetic field arrow is plotted in red when the ambiguity is resolved by applying the polarity method. The magnetic field arrow is plotted in orange when the ambiguity is resolved by applying the chirality method. The blue circle embraces parallel filament lines with alternating long-axis magnetic field components, as in Fig. 5 of \citet{Leroy-etal-83}.}
\label{1636}
\end{figure*}

\begin{figure*}
\resizebox{\hsize}{!}{\includegraphics{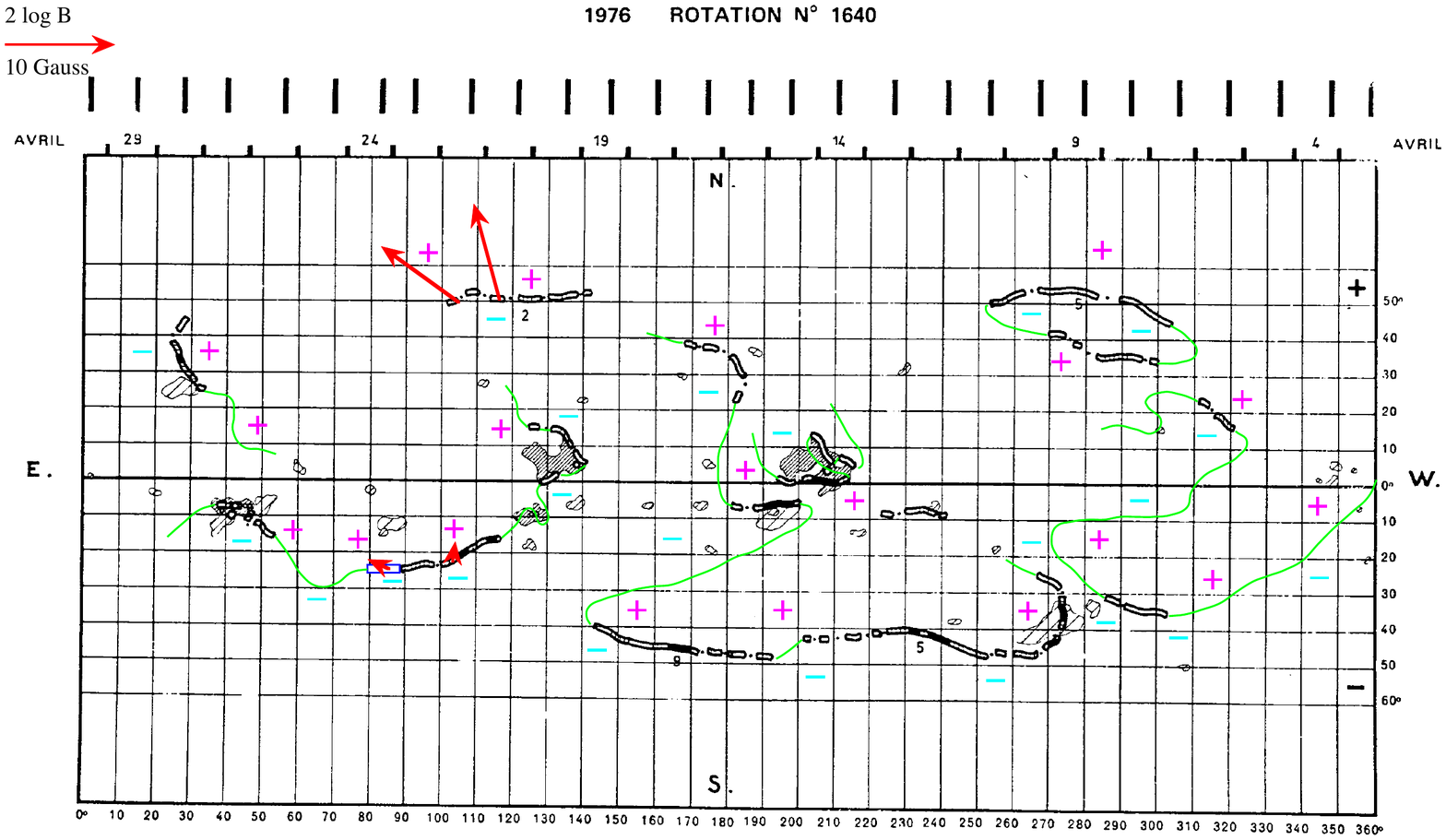}}
\caption{Synoptic map of rotation 1640. Horizontal axis: Carrington longitude. Vertical axis: Solar latitude. The neutral lines of the photospheric magnetic field taken from the McIntosh maps are added as green lines, and the polarities of the photospheric magnetic field also taken from the McIntosh maps are referred to with plus (for positive polarity, in purple) and minus (for negative polarity, in cyan) signs on either side of the neutral lines. Each measured magnetic field vector (average on an observed prominence) is represented by a vector of length $2\mathrm{log}B$. This length for $B=10$ Gauss is plotted in the upper left corner. The magnetic field arrow is plotted in red when the ambiguity is resolved by applying the polarity method. The magnetic field arrow is plotted in orange when the ambiguity is resolved by applying the chirality method.}
\label{1640}
\end{figure*}

\begin{figure*}
\resizebox{\hsize}{!}{\includegraphics{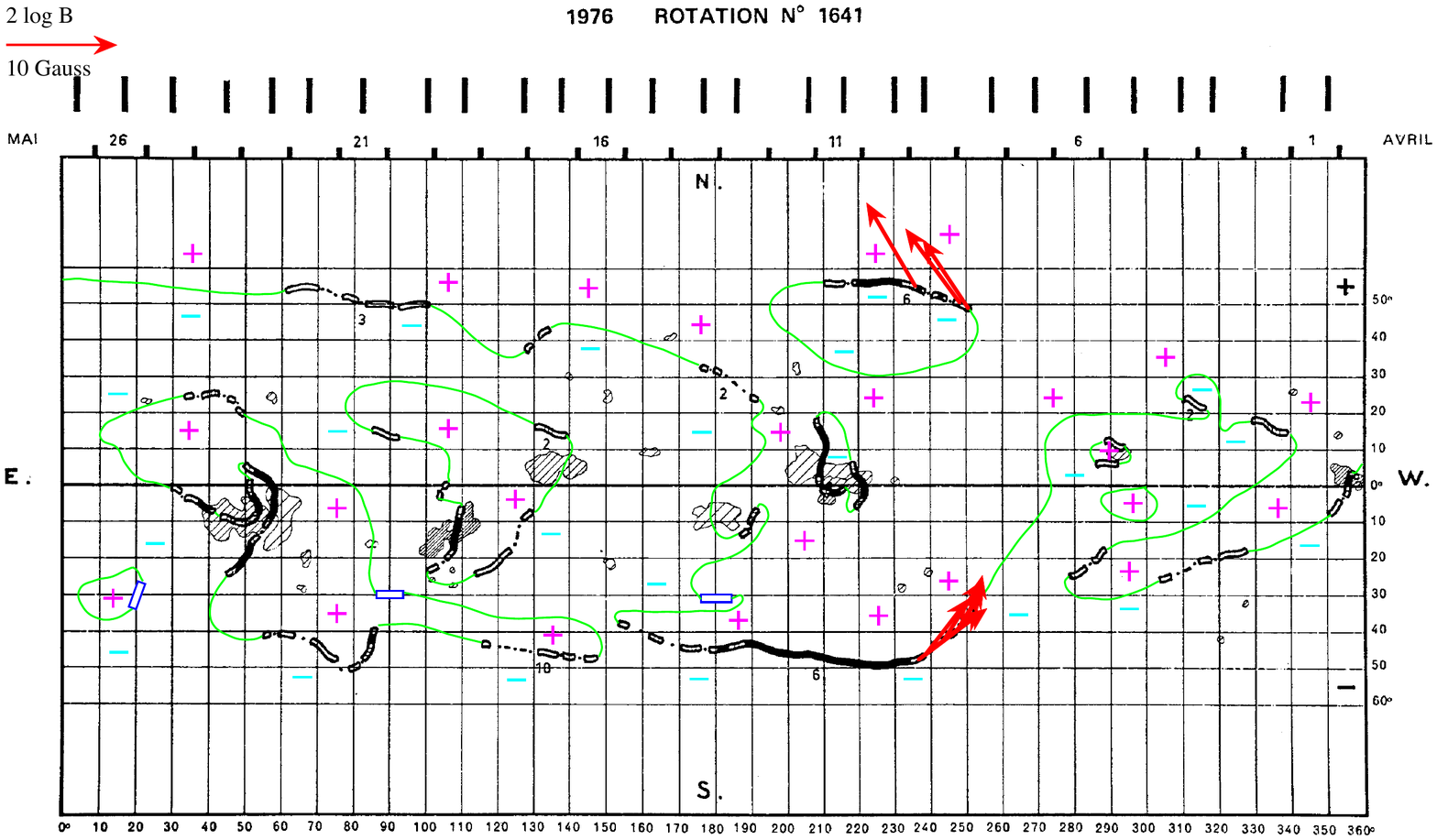}}
\caption{Synoptic map of rotation 1641. Horizontal axis: Carrington longitude. Vertical axis: Solar latitude. The neutral lines of the photospheric magnetic field taken from the McIntosh maps are added as green lines, and the polarities of the photospheric magnetic field also taken from the McIntosh maps are referred to with plus (for positive polarity, in purple) and minus (for negative polarity, in cyan) signs on either side of the neutral lines. Each measured magnetic field vector (average on an observed prominence) is represented by a vector of length $2\mathrm{log}B$. This length for $B=10$ Gauss is plotted in the upper left corner. The magnetic field arrow is plotted in red when the ambiguity is resolved by applying the polarity method. The magnetic field arrow is plotted in orange when the ambiguity is resolved by applying the chirality method.}
\label{1641}
\end{figure*}

\begin{figure*}
\resizebox{\hsize}{!}{\includegraphics{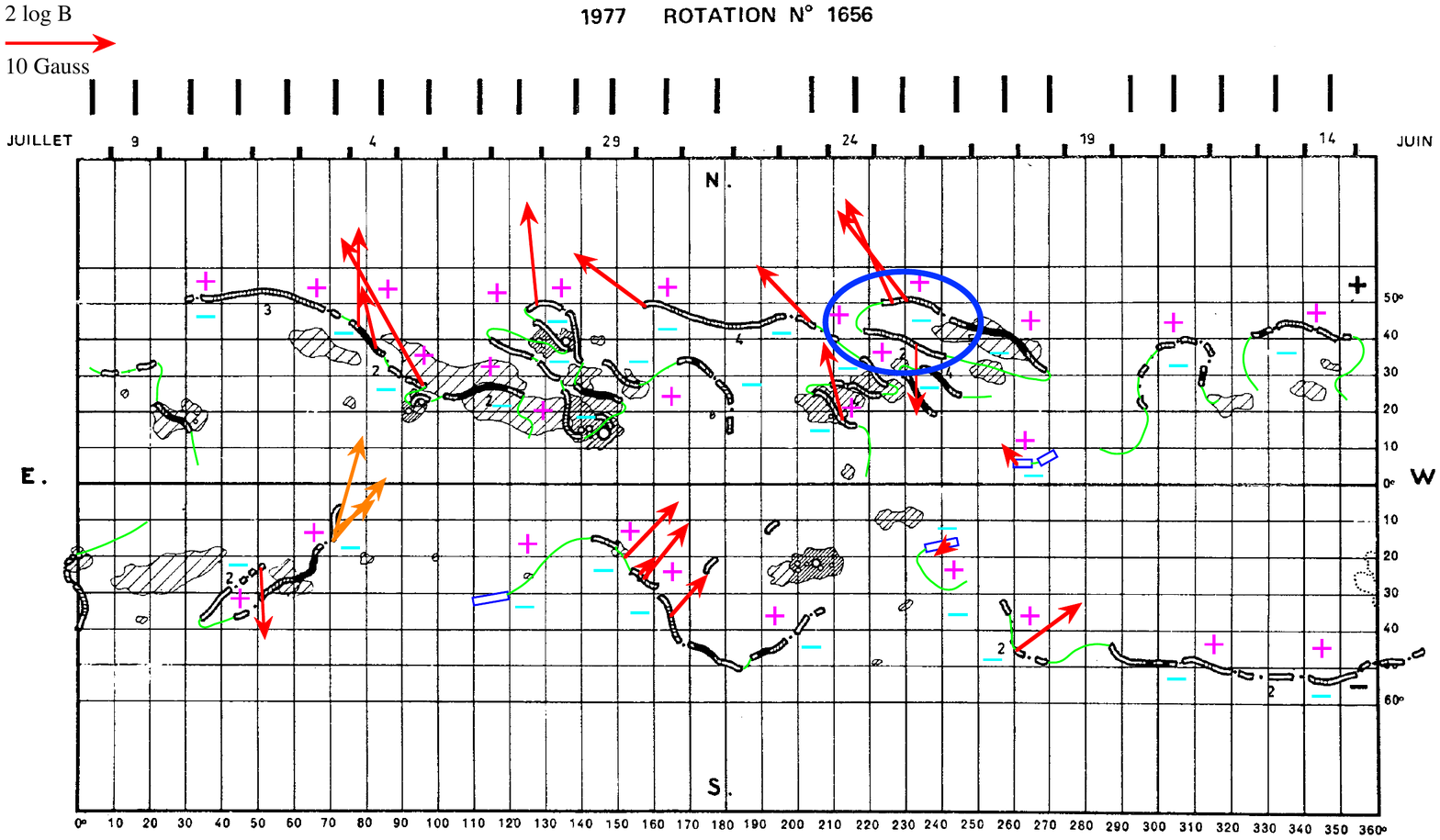}}
\caption{Synoptic map of rotation 1656. Horizontal axis: Carrington longitude. Vertical axis: Solar latitude. The neutral lines of the photospheric magnetic field taken from the McIntosh maps are added as green lines, and the polarities of the photospheric magnetic field also taken from the McIntosh maps are referred to with plus (for positive polarity, in purple) and minus (for negative polarity, in cyan) signs on either side of the neutral lines. Each measured magnetic field vector (average on an observed prominence) is represented by a vector of length $2\mathrm{log}B$. This length for $B=10$ Gauss is plotted in the upper left corner. The magnetic field arrow is plotted in red when the ambiguity is resolved by applying the polarity method. The magnetic field arrow is plotted in orange when the ambiguity is resolved by applying the chirality method. The blue circle embraces parallel filament lines with alternating long-axis magnetic field components, as in Fig. 5 of \citet{Leroy-etal-83}.}
\label{1656}
\end{figure*}

\begin{figure*}
\resizebox{\hsize}{!}{\includegraphics{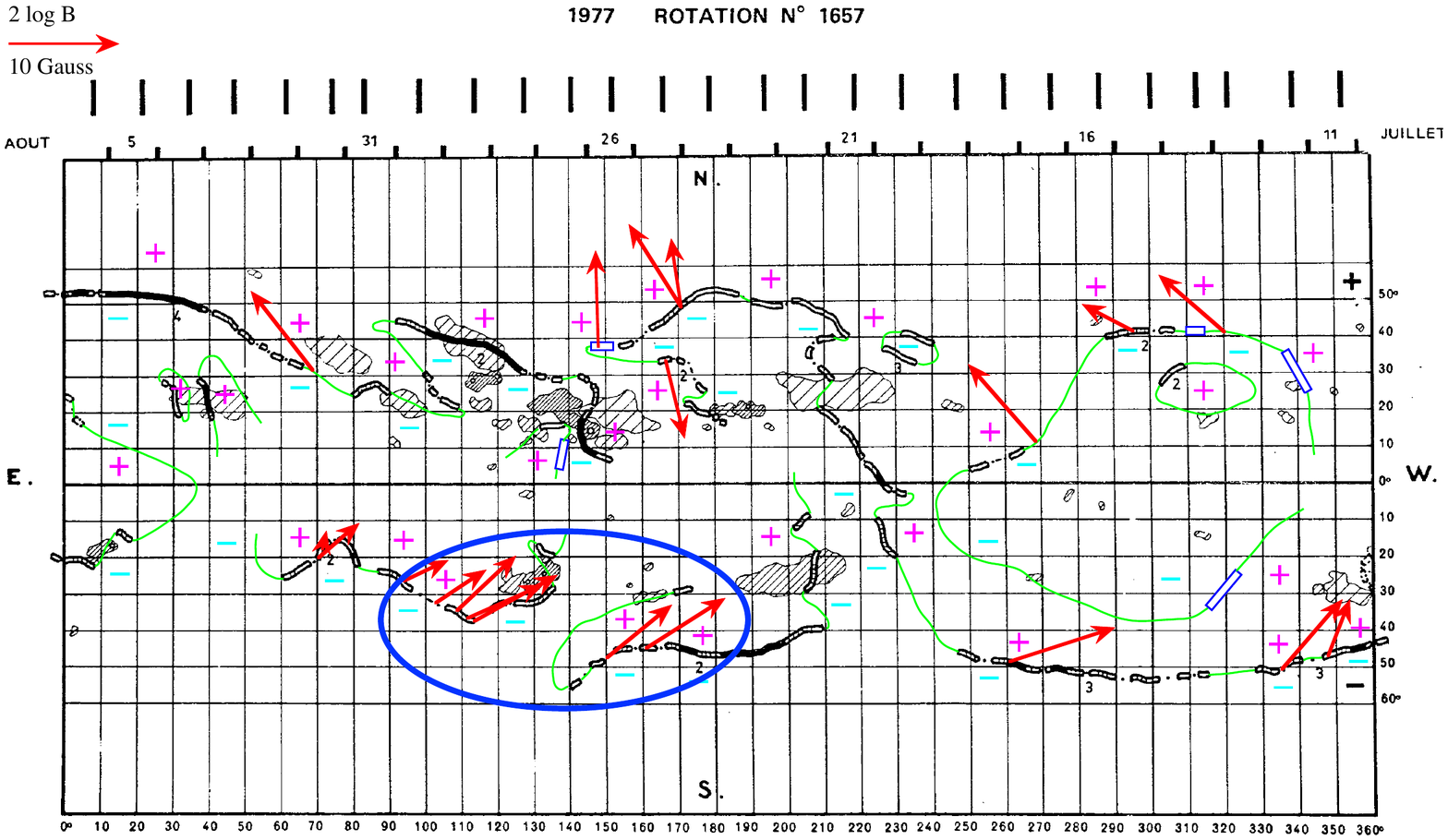}}
\caption{Synoptic map of rotation 1657. Horizontal axis: Carrington longitude. Vertical axis: Solar latitude. The neutral lines of the photospheric magnetic field taken from the McIntosh maps are added as green lines, and the polarities of the photospheric magnetic field also taken from the McIntosh maps are referred to with plus (for positive polarity, in purple) and minus (for negative polarity, in cyan) signs on either side of the neutral lines. Each measured magnetic field vector (average on an observed prominence) is represented by a vector of length $2\mathrm{log}B$. This length for $B=10$ Gauss is plotted in the upper left corner. The magnetic field arrow is plotted in red when the ambiguity is resolved by applying the polarity method. The magnetic field arrow is plotted in orange when the ambiguity is resolved by applying the chirality method. The blue circle embraces parallel filament lines (with an in-between neutral line) with alternating long-axis magnetic field components, as in Fig. 5 of \citet{Leroy-etal-83}.}
\label{1657}
\end{figure*}

\begin{figure*}
\resizebox{\hsize}{!}{\includegraphics{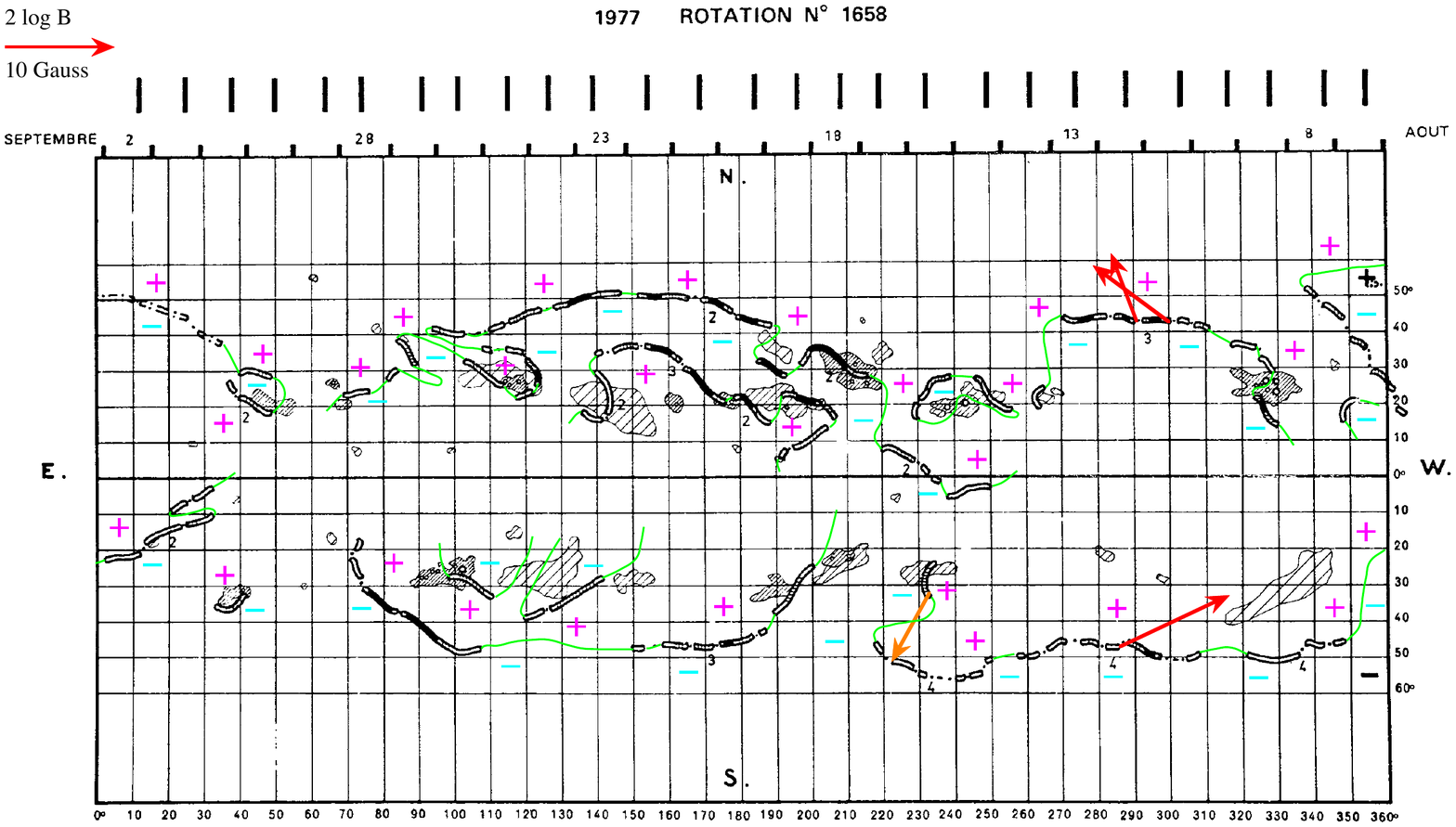}}
\caption{Synoptic map of rotation 1658. Horizontal axis: Carrington longitude. Vertical axis: Solar latitude. The neutral lines of the photospheric magnetic field taken from the McIntosh maps are added as green lines, and the polarities of the photospheric magnetic field also taken from the McIntosh maps are referred to with plus (for positive polarity, in purple) and minus (for negative polarity, in cyan) signs on either side of the neutral lines. Each measured magnetic field vector (average on an observed prominence) is represented by a vector of length $2\mathrm{log}B$. This length for $B=10$ Gauss is plotted in the upper left corner. The magnetic field arrow is plotted in red when the ambiguity is resolved by applying the polarity method. The magnetic field arrow is plotted in orange when the ambiguity is resolved by applying the chirality method.}
\label{1658}
\end{figure*}

\begin{figure*}
\resizebox{\hsize}{!}{\includegraphics{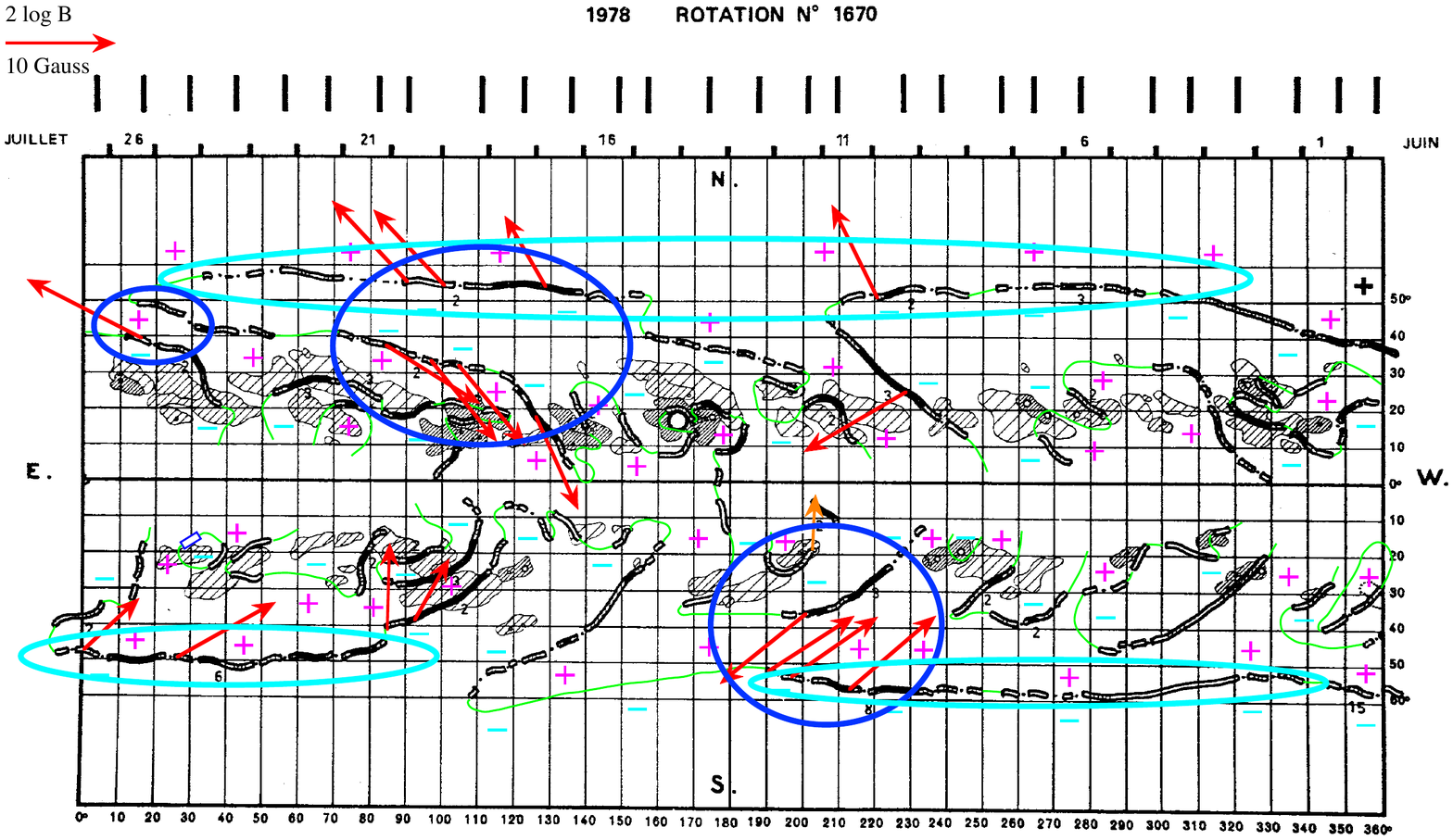}}
\caption{Synoptic map of rotation 1670. Horizontal axis: Carrington longitude. Vertical axis: Solar latitude. The neutral lines of the photospheric magnetic field taken from the McIntosh maps are added as green lines, and the polarities of the photospheric magnetic field also taken from the McIntosh maps are referred to with plus (for positive polarity, in purple) and minus (for negative polarity, in cyan) signs on either side of the neutral lines. Each measured magnetic field vector (average on an observed prominence) is represented by a vector of length $2\mathrm{log}B$. This length for $B=10$ Gauss is plotted in the upper left corner. The magnetic field arrow is plotted in red when the ambiguity is resolved by applying the polarity method. The magnetic field arrow is plotted in orange when the ambiguity is resolved by applying the chirality method. The blue circles embrace parallel filament lines with alternating long-axis magnetic field components, as in Fig. 5 of \citet{Leroy-etal-83}. The cyan ellipses embrace the Polar Crown in formation.}
\label{1670}
\end{figure*}

\begin{figure*}
\resizebox{\hsize}{!}{\includegraphics{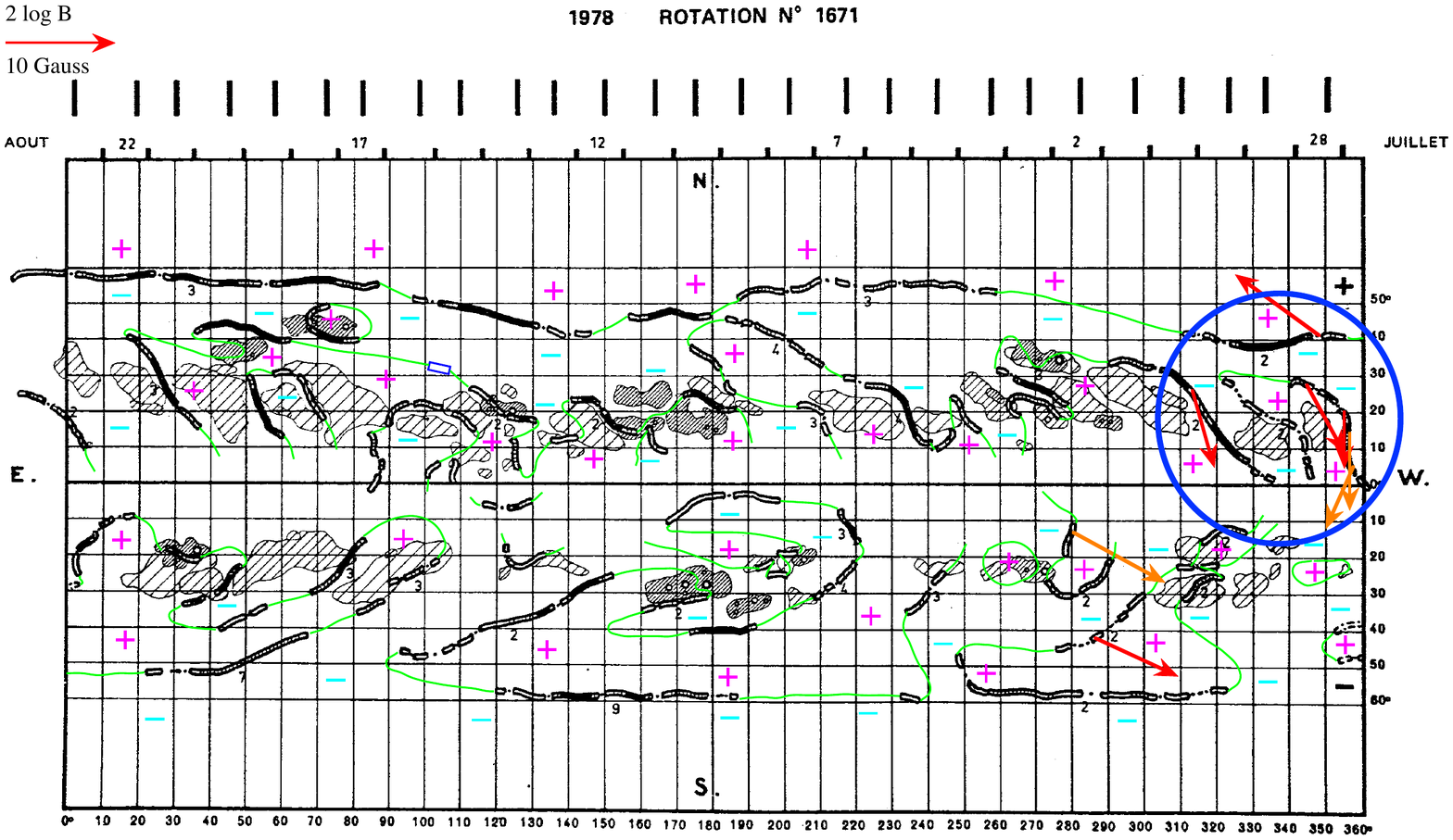}}
\caption{Synoptic map of rotation 1671. Horizontal axis: Carrington longitude. Vertical axis: Solar latitude. The neutral lines of the photospheric magnetic field taken from the McIntosh maps are added as green lines, and the polarities of the photospheric magnetic field also taken from the McIntosh maps are referred to with plus (for positive polarity, in purple) and minus (for negative polarity, in cyan) signs on either side of the neutral lines. Each measured magnetic field vector (average on an observed prominence) is represented by a vector of length $2\mathrm{log}B$. This length for $B=10$ Gauss is plotted in the upper left corner. The magnetic field arrow is plotted in red when the ambiguity is resolved by applying the polarity method. The magnetic field arrow is plotted in orange when the ambiguity is resolved by applying the chirality method. The blue circle embraces parallel filament lines (with an in-bteween filament line) with alternating long-axis magnetic field components, as in Fig. 5 of \citet{Leroy-etal-83}.}
\label{1671}
\end{figure*}

\begin{figure*}
\resizebox{\hsize}{!}{\includegraphics{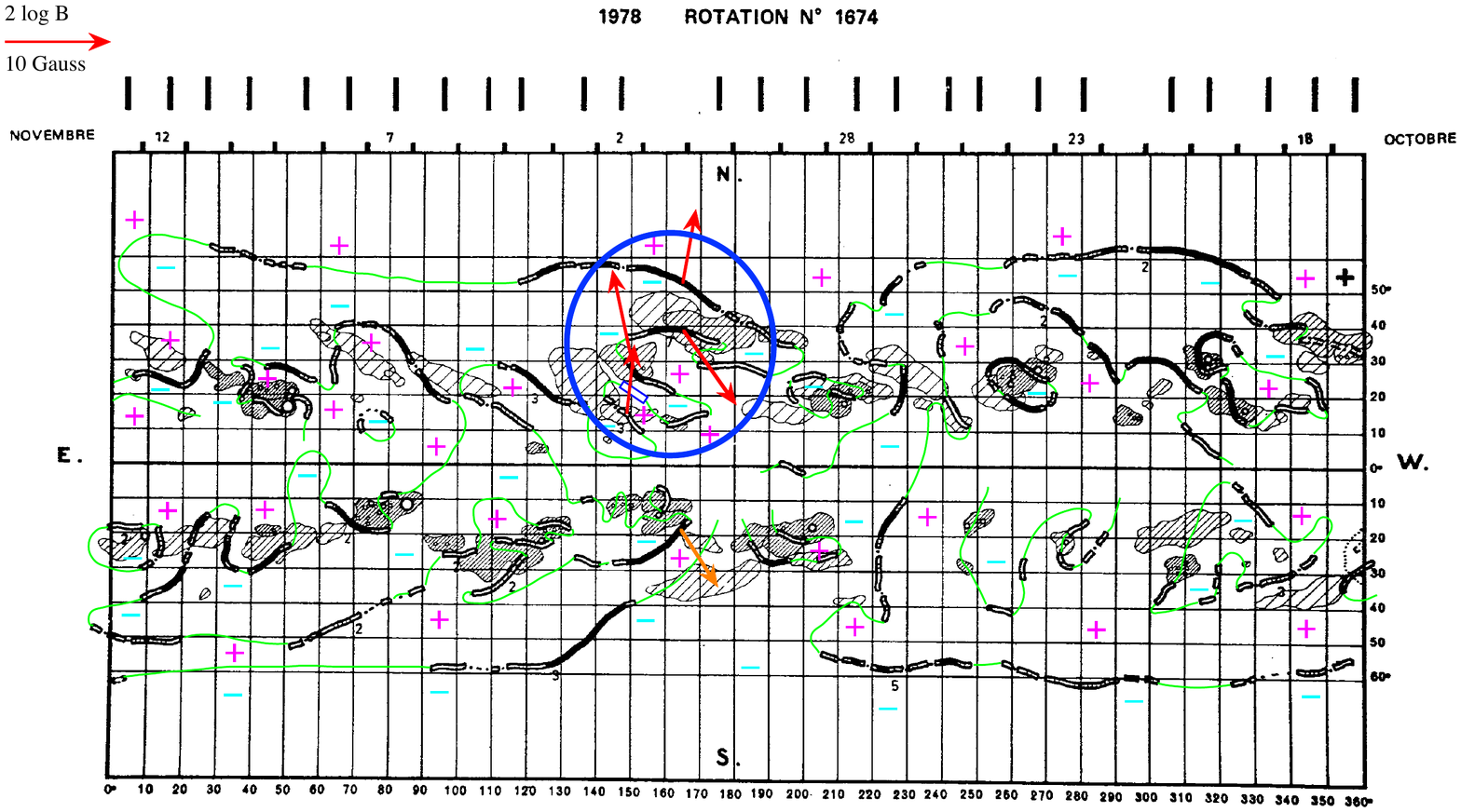}}
\caption{Synoptic map of rotation 1674.Horizontal axis: Carrington longitude. Vertical axis: Solar latitude. The neutral lines of the photospheric magnetic field taken from the McIntosh maps are added as green lines, and the polarities of the photospheric magnetic field also taken from the McIntosh maps are referred to with plus (for positive polarity, in purple) and minus (for negative polarity, in cyan) signs on either side of the neutral lines. Each measured magnetic field vector (average on an observed prominence) is represented by a vector of length $2\mathrm{log}B$. This length for $B=10$ Gauss is plotted in the upper left corner. The magnetic field arrow is plotted in red when the ambiguity is resolved by applying the polarity method. The magnetic field arrow is plotted in orange when the ambiguity is resolved by applying the chirality method. The blue circle embraces parallel filament lines with alternating long-axis magnetic field components, as in Fig. 5 of \citet{Leroy-etal-83}.}
\label{1674}
\end{figure*}

\begin{figure*}
\resizebox{\hsize}{!}{\includegraphics{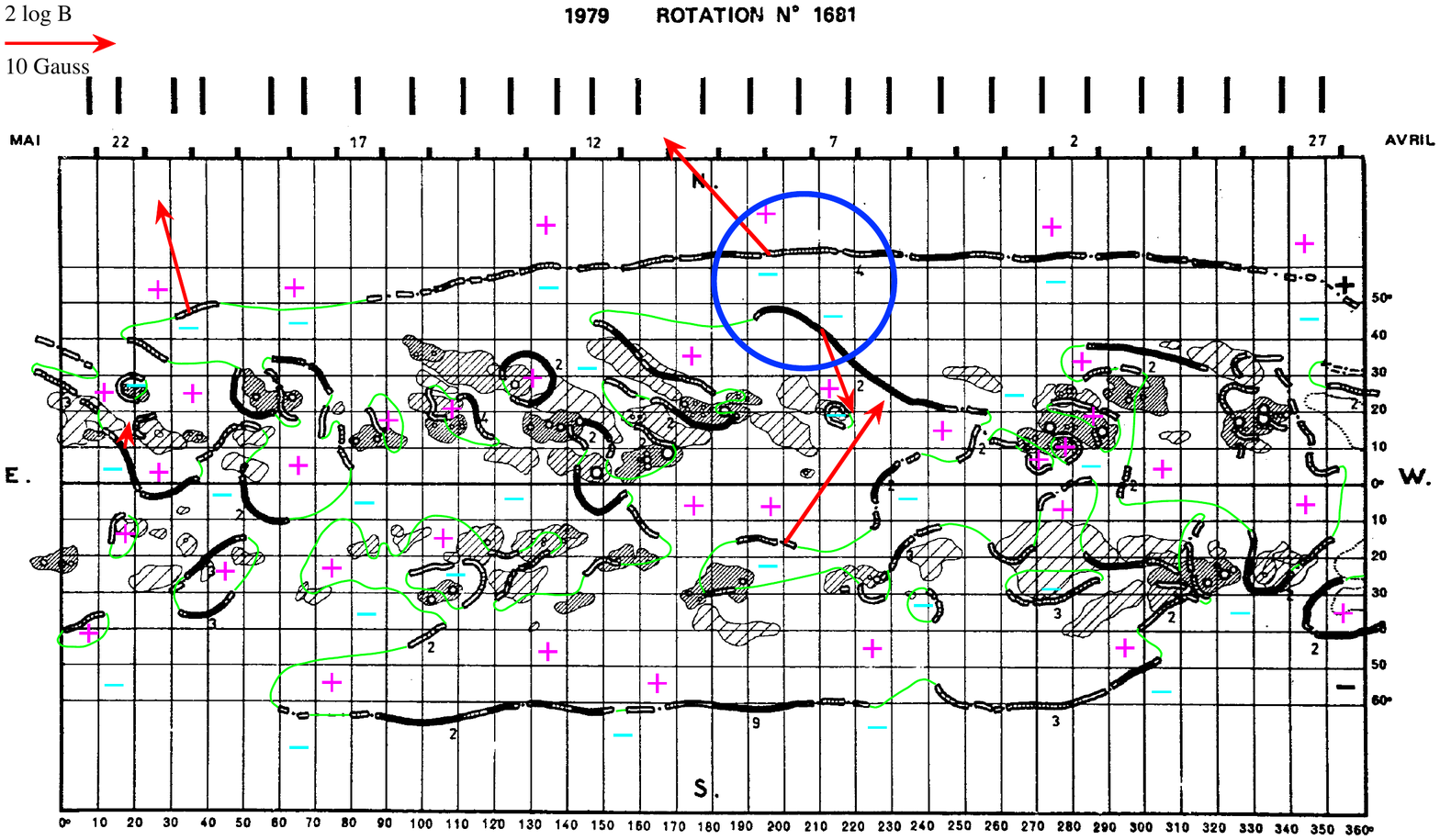}}
\caption{Synoptic map of rotation 1681. Horizontal axis: Carrington longitude. Vertical axis: Solar latitude. The neutral lines of the photospheric magnetic field taken from the McIntosh maps are added as green lines, and the polarities of the photospheric magnetic field also taken from the McIntosh maps are referred to with plus (for positive polarity, in purple) and minus (for negative polarity, in cyan) signs on either side of the neutral lines. Each measured magnetic field vector (average on an observed prominence) is represented by a vector of length $2\mathrm{log}B$. This length for $B=10$ Gauss is plotted in the upper left corner. The magnetic field arrow is plotted in red when the ambiguity is resolved by applying the polarity method. The magnetic field arrow is plotted in orange when the ambiguity is resolved by applying the chirality method. The blue circle embraces parallel filament lines with alternating long-axis magnetic field components, as in Fig. 5 of \citet{Leroy-etal-83}.}
\label{1681}
\end{figure*}

\begin{figure*}
\resizebox{\hsize}{!}{\includegraphics{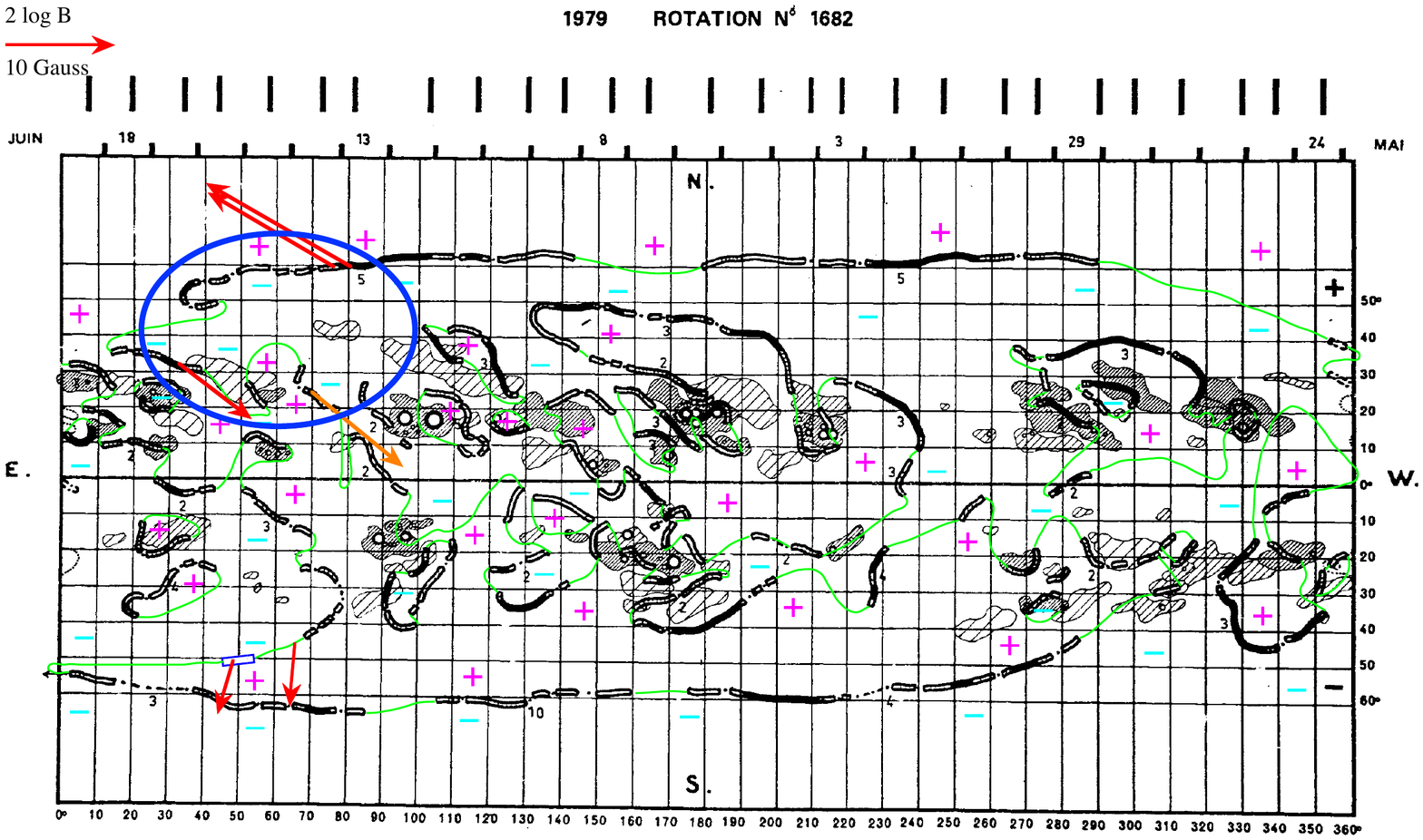}}
\caption{Synoptic map of rotation 1682. Horizontal axis: Carrington longitude. Vertical axis: Solar latitude. The neutral lines of the photospheric magnetic field taken from the McIntosh maps are added as green lines, and the polarities of the photospheric magnetic field also taken from the McIntosh maps are referred to with plus (for positive polarity, in purple) and minus (for negative polarity, in cyan) signs on either side of the neutral lines. Each measured magnetic field vector (average on an observed prominence) is represented by a vector of length $2\mathrm{log}B$. This length for $B=10$ Gauss is plotted in the upper left corner. The magnetic field arrow is plotted in red when the ambiguity is resolved by applying the polarity method. The magnetic field arrow is plotted in orange when the ambiguity is resolved by applying the chirality method. The blue circle embraces parallel filament lines with alternating long-axis magnetic field components, as in Fig. 5 of \citet{Leroy-etal-83}.}
\label{1682}
\end{figure*}

\begin{figure*}
\resizebox{\hsize}{!}{\includegraphics{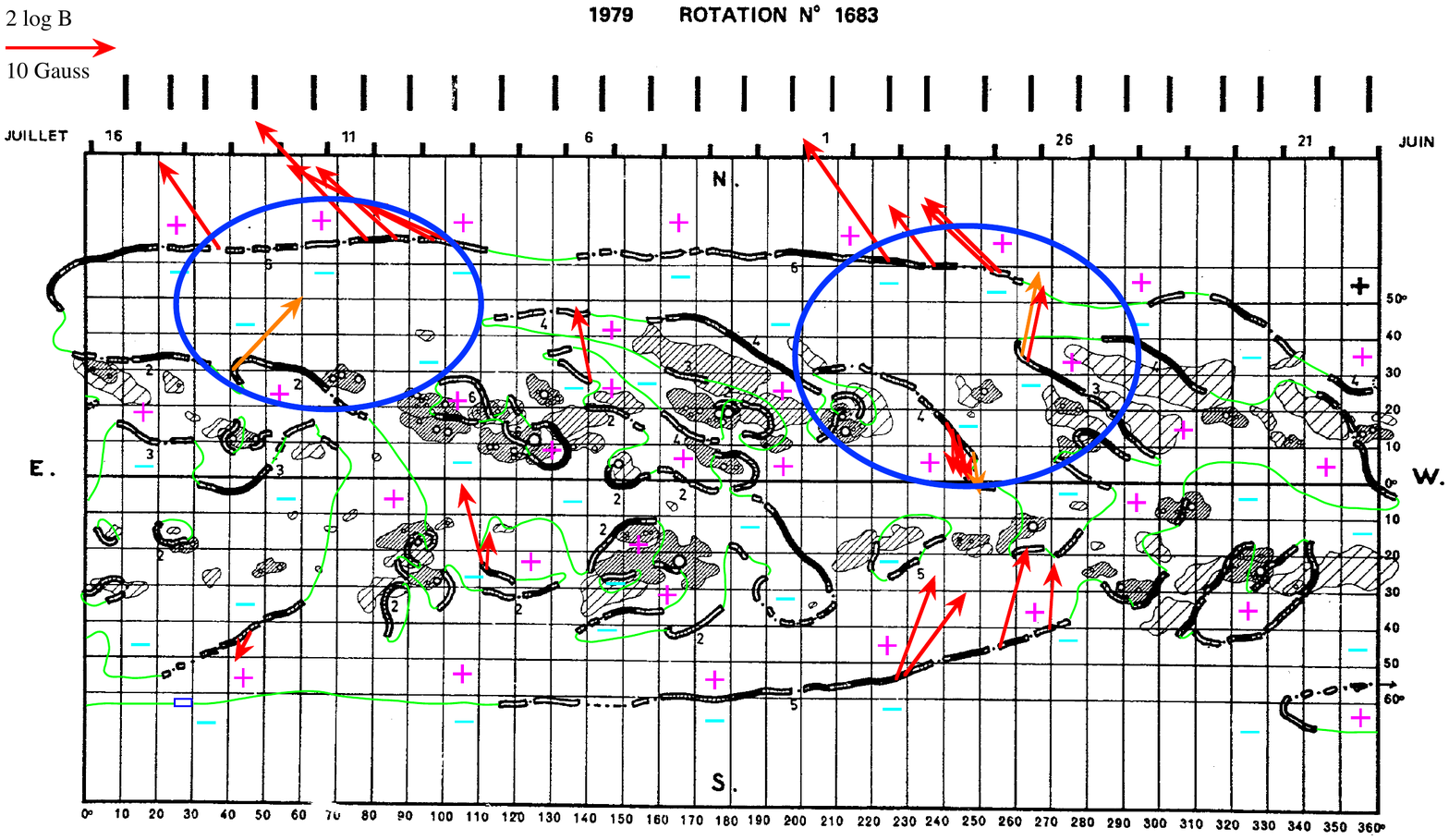}}
\caption{Synoptic map of rotation 1683. Horizontal axis: Carrington longitude. Vertical axis: Solar latitude. The neutral lines of the photospheric magnetic field taken from the McIntosh maps are added as green lines, and the polarities of the photospheric magnetic field also taken from the McIntosh maps are referred to with plus (for positive polarity, in purple) and minus (for negative polarity, in cyan) signs on either side of the neutral lines. Each measured magnetic field vector (average on an observed prominence) is represented by a vector of length $2\mathrm{log}B$. This length for $B=10$ Gauss is plotted in the upper left corner. The magnetic field arrow is plotted in red when the ambiguity is resolved by applying the polarity method. The magnetic field arrow is plotted in orange when the ambiguity is resolved by applying the chirality method. The blue circles embrace parallel filament lines with alternating long-axis magnetic field components, as in Fig. 5 of \citet{Leroy-etal-83}.}
\label{1683}
\end{figure*}

\begin{figure*}
\resizebox{\hsize}{!}{\includegraphics{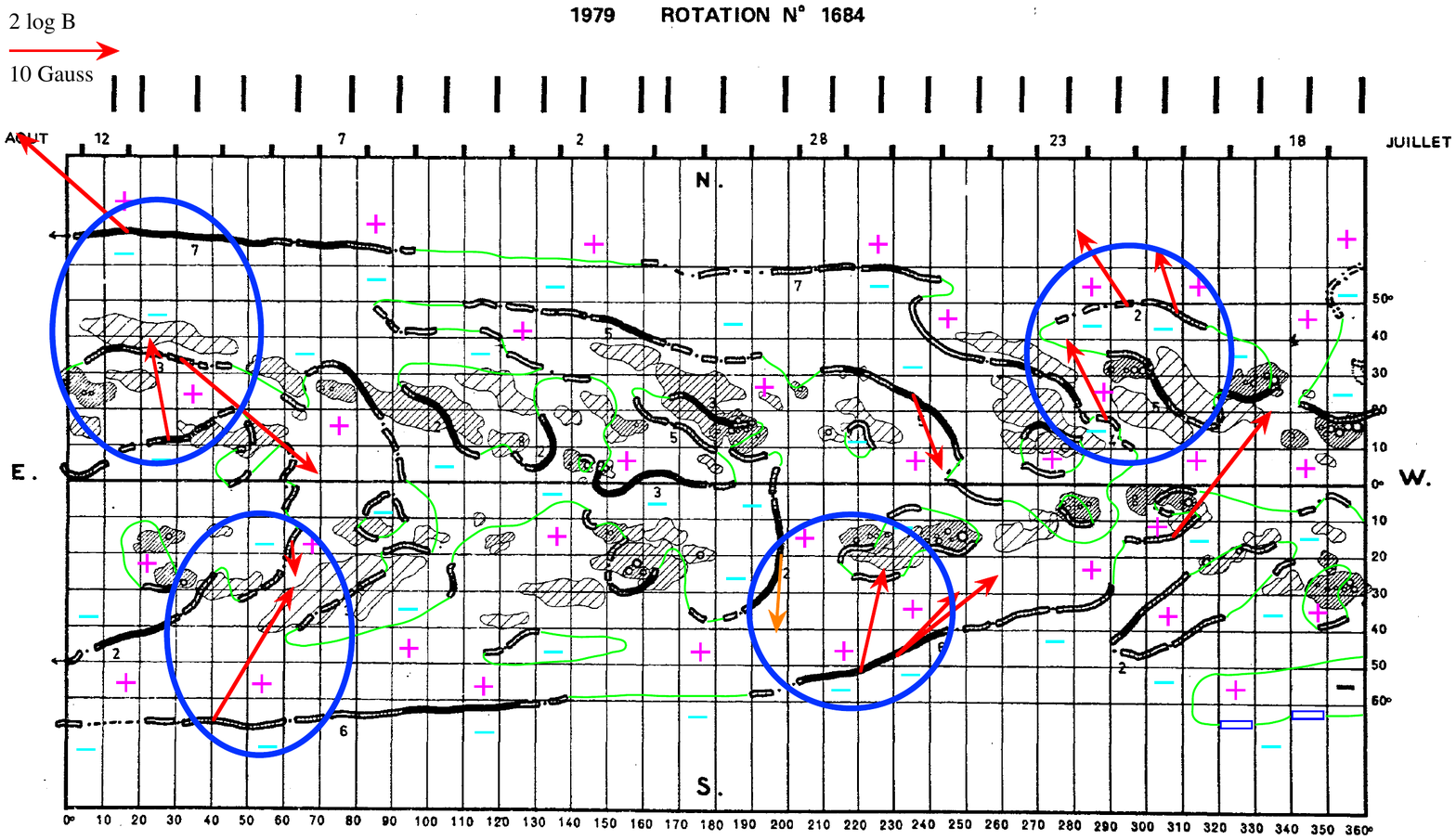}}
\caption{Synoptic map of rotation 1684. Horizontal axis: Carrington longitude. Vertical axis: Solar latitude. The neutral lines of the photospheric magnetic field taken from the McIntosh maps are added as green lines, and the polarities of the photospheric magnetic field also taken from the McIntosh maps are referred to with plus (for positive polarity, in purple) and minus (for negative polarity, in cyan) signs on either side of the neutral lines. Each measured magnetic field vector (average on an observed prominence) is represented by a vector of length $2\mathrm{log}B$. This length for $B=10$ Gauss is plotted in the upper left corner. The magnetic field arrow is plotted in red when the ambiguity is resolved by applying the polarity method. The magnetic field arrow is plotted in orange when the ambiguity is resolved by applying the chirality method. The blue circles embrace parallel filament lines with alternating long-axis magnetic field components, as in Fig. 5 of \citet{Leroy-etal-83}. For one of them, there is an in-between alternating filament line.}
\label{1684}
\end{figure*}

\begin{figure*}
\resizebox{\hsize}{!}{\includegraphics{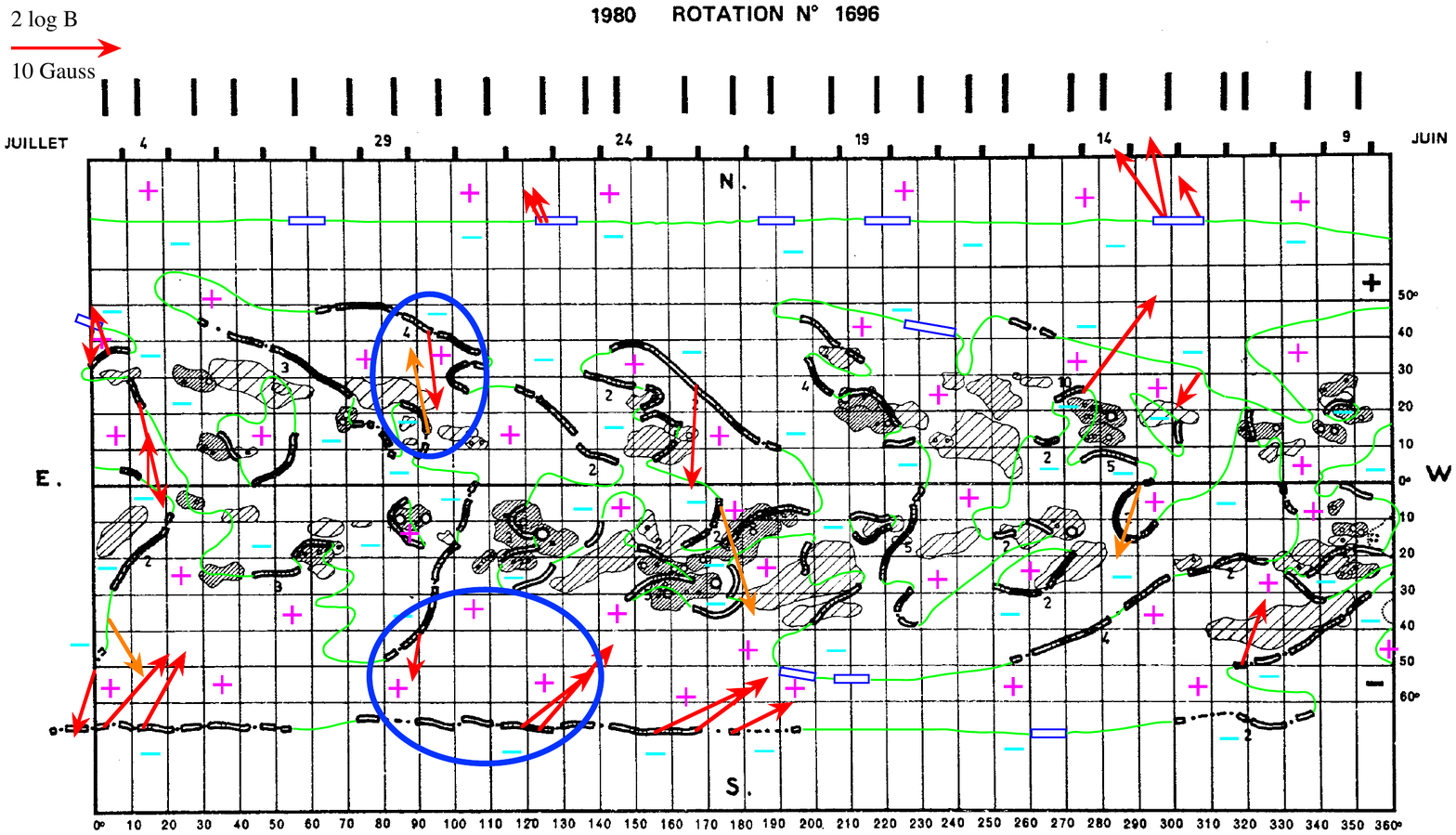}}
\caption{Synoptic map of rotation 1696. Horizontal axis: Carrington longitude. Vertical axis: Solar latitude. The neutral lines of the photospheric magnetic field taken from the McIntosh maps are added as green lines, and the polarities of the photospheric magnetic field also taken from the McIntosh maps are referred to with plus (for positive polarity, in purple) and minus (for negative polarity, in cyan) signs on either side of the neutral lines. Each measured magnetic field vector (average on an observed prominence) is represented by a vector of length $2\mathrm{log}B$. This length for $B=10$ Gauss is plotted in the upper left corner. The magnetic field arrow is plotted in red when the ambiguity is resolved by applying the polarity method. The magnetic field arrow is plotted in orange when the ambiguity is resolved by applying the chirality method. The blue circles embrace parallel filament lines with alternating long-axis magnetic field components, as in Fig. 5 of \citet{Leroy-etal-83}.}
\label{1696}
\end{figure*}

\begin{figure*}
\resizebox{\hsize}{!}{\includegraphics{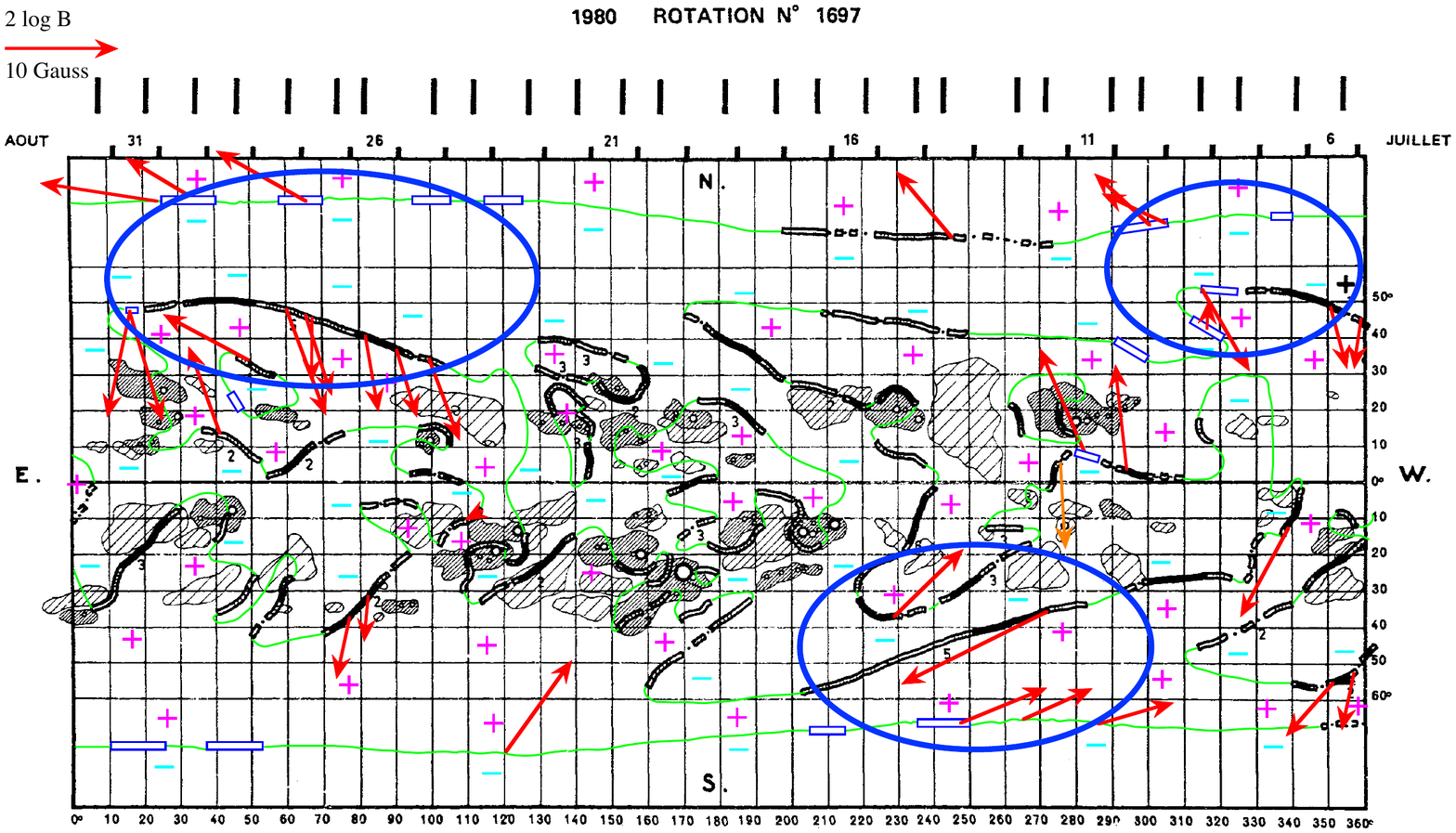}}
\caption{Synoptic map of rotation 1697. Horizontal axis: Carrington longitude. Vertical axis: Solar latitude. The neutral lines of the photospheric magnetic field taken from the McIntosh maps are added as green lines, and the polarities of the photospheric magnetic field also taken from the McIntosh maps are referred to with plus (for positive polarity, in purple) and minus (for negative polarity, in cyan) signs on either side of the neutral lines. Each measured magnetic field vector (average on an observed prominence) is represented by a vector of length $2\mathrm{log}B$. This length for $B=10$ Gauss is plotted in the upper left corner. The magnetic field arrow is plotted in red when the ambiguity is resolved by applying the polarity method. The magnetic field arrow is plotted in orange when the ambiguity is resolved by applying the chirality method. The blue circles embrace parallel filament lines with alternating long-axis magnetic field components, as in Fig. 5 of \citet{Leroy-etal-83}.}
\label{1697}
\end{figure*}

\begin{figure*}
\resizebox{\hsize}{!}{\includegraphics{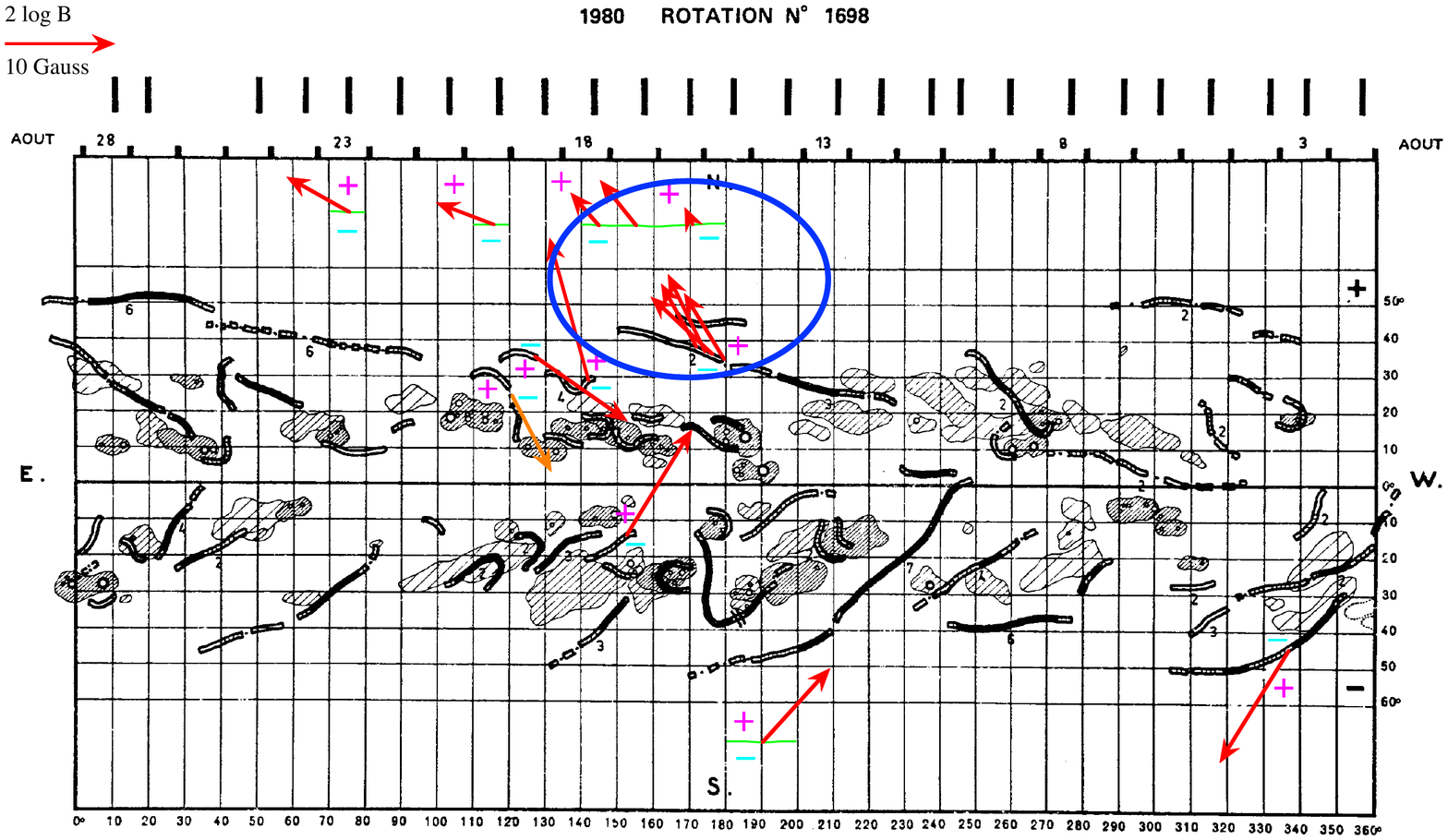}}
\caption{Synoptic map of rotation 1698}. Horizontal axis: Carrington longitude. Vertical axis: Solar latitude. The neutral lines of the photospheric magnetic field taken from the McIntosh maps are added as green lines, and the polarities of the photospheric magnetic field also taken from the McIntosh maps are referred to with plus (for positive polarity, in purple) and minus (for negative polarity, in cyan) signs on either side of the neutral lines. Each measured magnetic field vector (average on an observed prominence) is represented by a vector of length $2\mathrm{log}B$. This length for $B=10$ Gauss is plotted in the upper left corner. The magnetic field arrow is plotted in red when the ambiguity is resolved by applying the polarity method. The magnetic field arrow is plotted in orange when the ambiguity is resolved by applying the chirality method. The blue circle embraces parallel filament lines (with an in-between filament line) with alternating long-axis magnetic field components, as in Fig. 5 of \citet{Leroy-etal-83}.
\label{1698}
\end{figure*}

\begin{figure*}
\resizebox{\hsize}{!}{\includegraphics{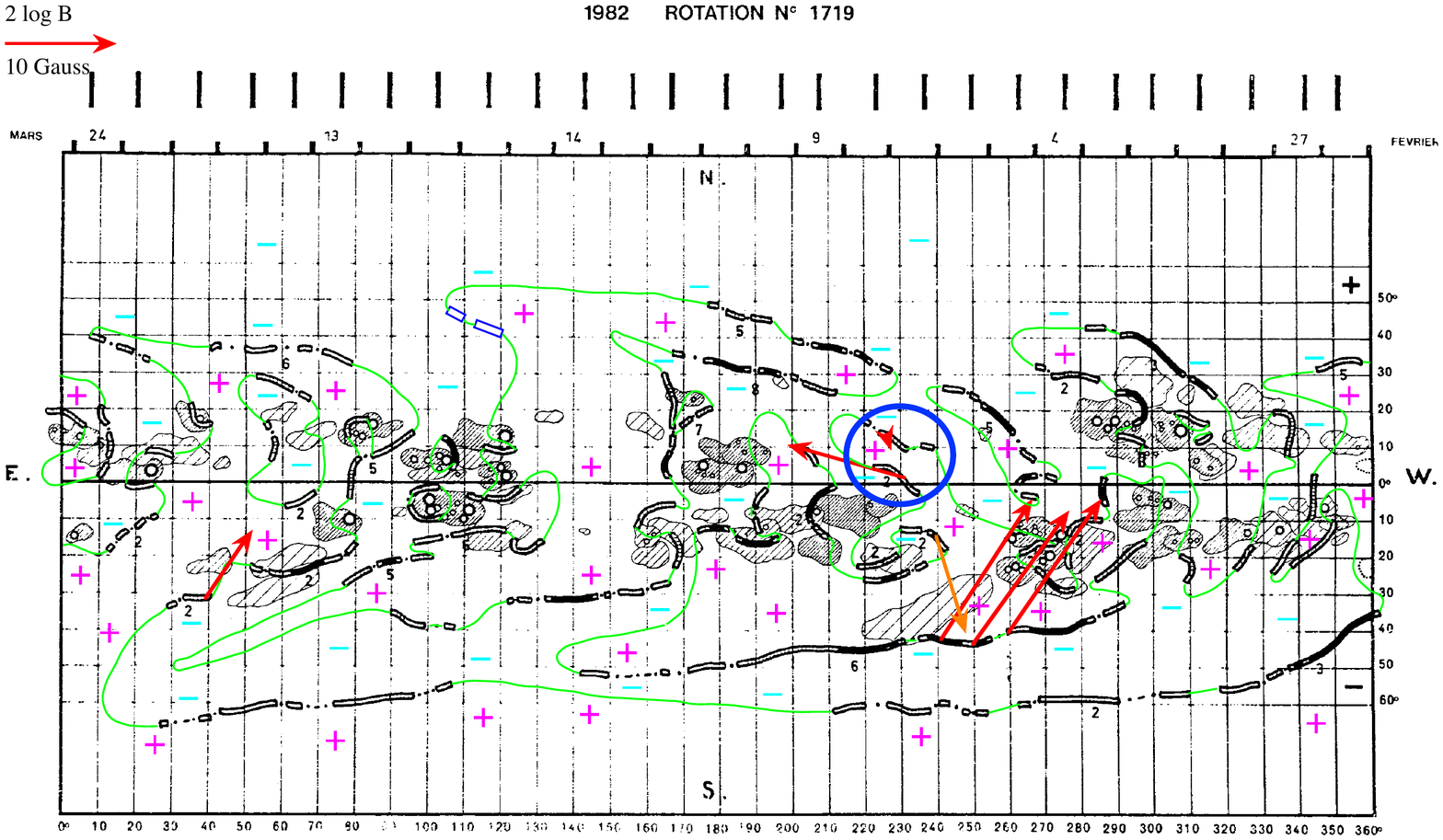}}
\caption{Synoptic map of rotation 1719. Horizontal axis: Carrington longitude. Vertical axis: Solar latitude. The neutral lines of the photospheric magnetic field taken from the McIntosh maps are added as green lines, and the polarities of the photospheric magnetic field also taken from the McIntosh maps are referred to with plus (for positive polarity, in purple) and minus (for negative polarity, in cyan) signs on either side of the neutral lines. Each measured magnetic field vector (average on an observed prominence) is represented by a vector of length $2\mathrm{log}B$. This length for $B=10$ Gauss is plotted in the upper left corner. The magnetic field arrow is plotted in red when the ambiguity is resolved by applying the polarity method. The magnetic field arrow is plotted in orange when the ambiguity is resolved by applying the chirality method. The blue circle embraces parallel filament lines with alternating long-axis magnetic field components, as in Fig. 5 of \citet{Leroy-etal-83}.}
\label{1719}
\end{figure*}

\begin{figure*}
\resizebox{\hsize}{!}{\includegraphics{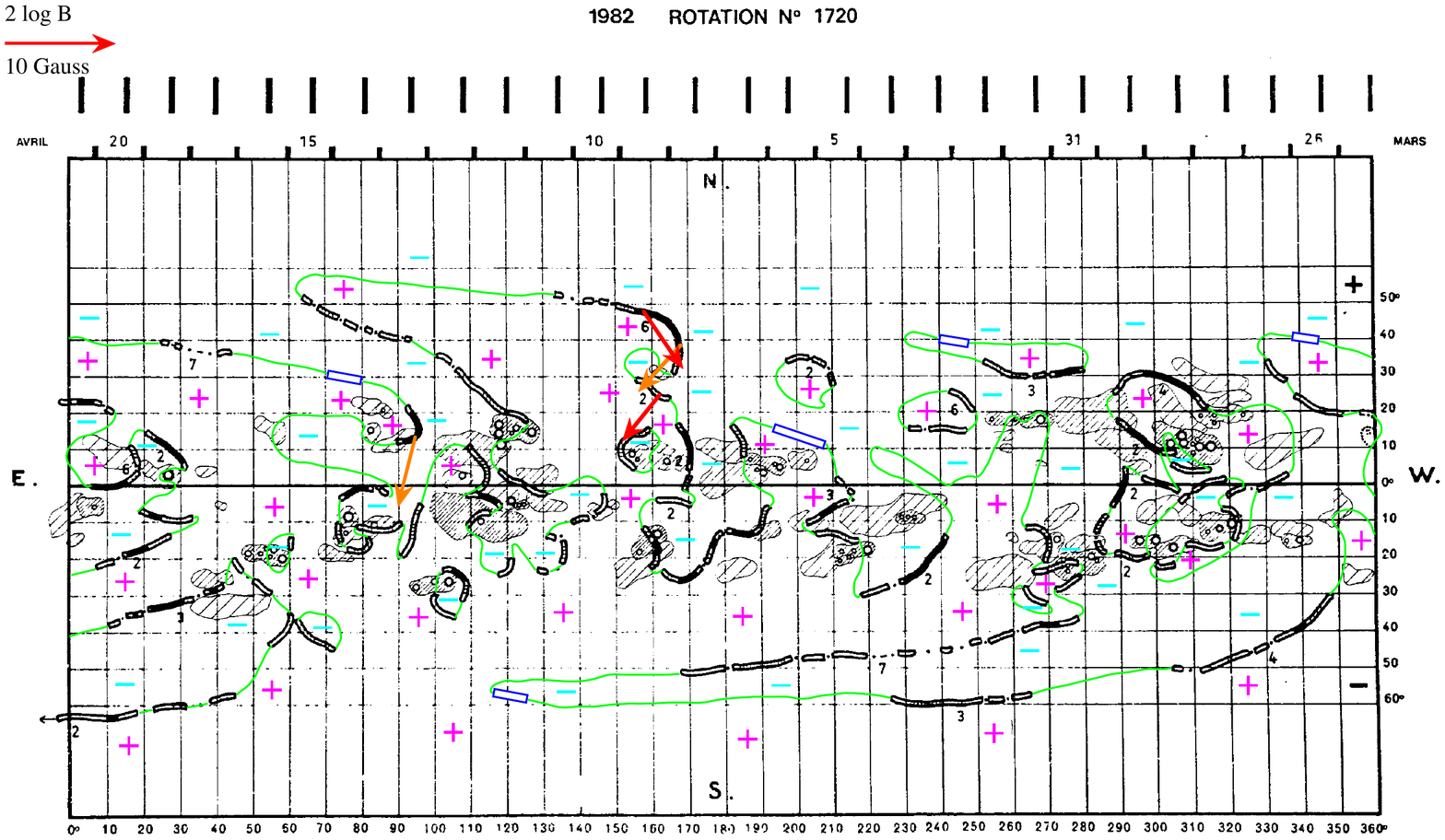}}
\caption{Synoptic map of rotation 1720. Horizontal axis: Carrington longitude. Vertical axis: Solar latitude. The neutral lines of the photospheric magnetic field taken from the McIntosh maps are added as green lines, and the polarities of the photospheric magnetic field also taken from the McIntosh maps are referred to with plus (for positive polarity, in purple) and minus (for negative polarity, in cyan) signs on either side of the neutral lines. Each measured magnetic field vector (average on an observed prominence) is represented by a vector of length $2\mathrm{log}B$. This length for $B=10$ Gauss is plotted in the upper left corner. The magnetic field arrow is plotted in red when the ambiguity is resolved by applying the polarity method. The magnetic field arrow is plotted in orange when the ambiguity is resolved by applying the chirality method.}
\label{1720}
\end{figure*}

\begin{figure*}
\resizebox{\hsize}{!}{\includegraphics{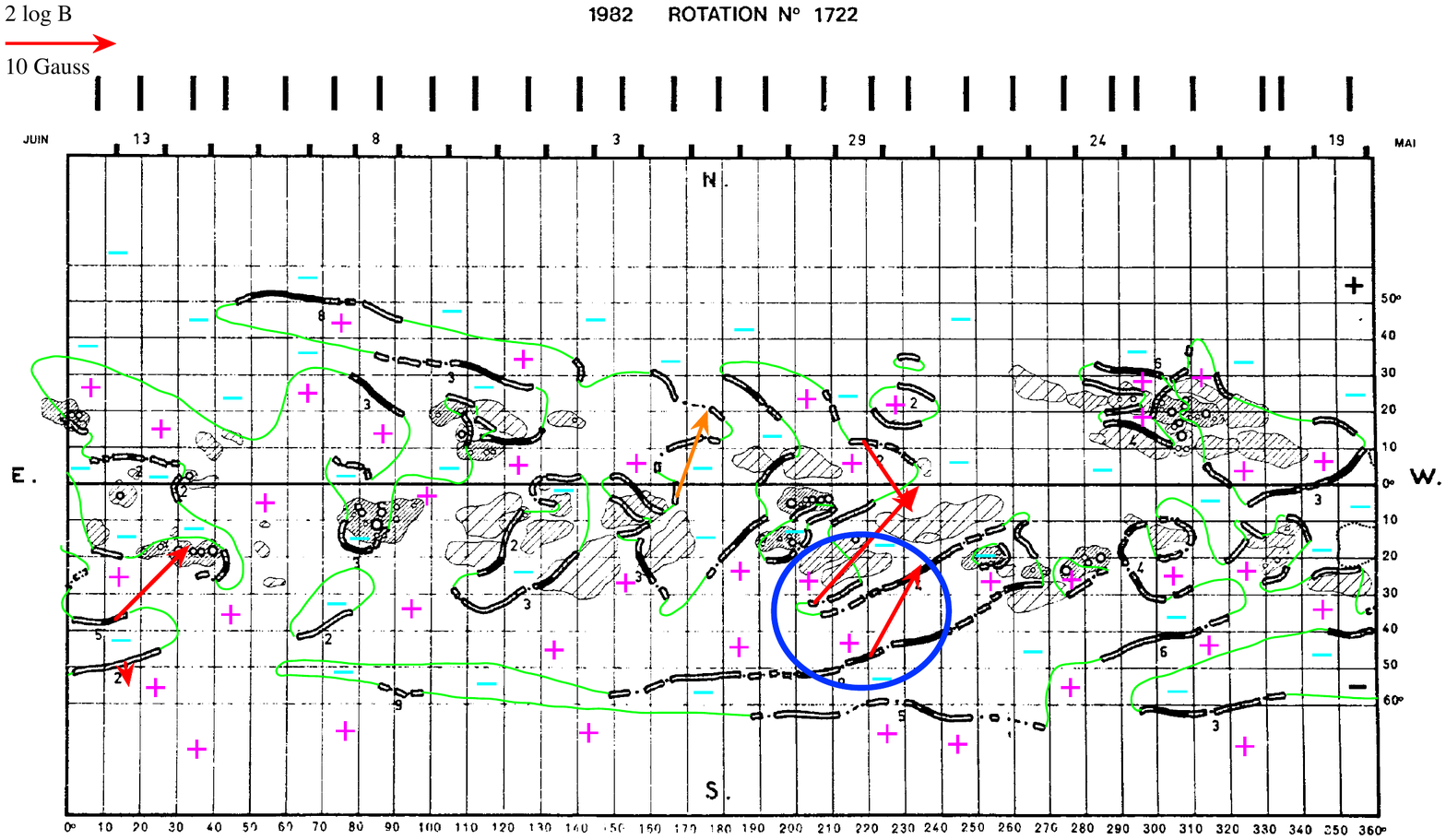}}
\caption{Synoptic map of rotation 1722. Horizontal axis: Carrington longitude. Vertical axis: Solar latitude. The neutral lines of the photospheric magnetic field taken from the McIntosh maps are added as green lines, and the polarities of the photospheric magnetic field also taken from the McIntosh maps are referred to with plus (for positive polarity, in purple) and minus (for negative polarity, in cyan) signs on either side of the neutral lines. Each measured magnetic field vector (average on an observed prominence) is represented by a vector of length $2\mathrm{log}B$. This length for $B=10$ Gauss is plotted in the upper left corner. The magnetic field arrow is plotted in red when the ambiguity is resolved by applying the polarity method. The magnetic field arrow is plotted in orange when the ambiguity is resolved by applying the chirality method. The blue circle embraces parallel filament lines (with an in-between filament line) with alternating long-axis magnetic field components, as in Fig. 5 of \citet{Leroy-etal-83}.}
\label{1722}
\end{figure*}

As a result, we obtained unambiguous average horizontal magnetic field vectors of 296 prominences, each associated with a filament observed eight days before (W limb) or after (E limb). We report these vectors in the synoptic maps in Figs. \ref{1616}-\ref{1722}. The magnetic field vector, one for each prominence, is reported as a red or orange arrow of length 2$\mathrm{log}B$, in order to show the weaker fields more clearly and to avoid too long arrows. The $B$ value, expressed in Gauss, has always been found to be higher than one. Thus, $\mathrm{log}B$ with $B$ in Gauss is always nonzero and positive. In the top-left corner of each plot, the red arrow represents $2\mathrm{log}B$ for $B=10$ Gauss and is provided as a reference unit. In the maps, the arrow is plotted in red when the ambiguity is resolved by applying the polarity method. The arrow is plotted in orange when the ambiguity is resolved by applying the chirality method.

Two main features become visible in these 24 maps. First, alternating field directions can be seen from one neutral line to the next. Numerous examples are outlined in Figs. \ref{1616}, \ref{1636}, \ref{1656}, \ref{1657}, \ref{1670}, \ref{1671}, \ref{1674}, \ref{1681}, \ref{1682}, \ref{1683}, \ref{1684}, \ref{1696}, \ref{1697}, \ref{1698}, \ref{1719}, and \ref{1722}. This alternation, which was represented in Fig. 5 of \citet{Leroy-etal-83}, is in fact a consequence or a proof of the chirality law.

Second, the field vectors are generally aligned with a solar north-south field distorted by the differential rotation effect. This differential rotation effect has been pointed out by \citet[][Fig. 2]{Hyder-65}, but with a sign error about the field direction. This is shown in a statistical way in Fig. \ref{azimuths}, where the angle between the field vector and the solar parallel (modulo 180$^{\circ}$ to account for the alternation cited above) is linearly fit for the 296 prominences as a function of the solar latitude. The 90$^{\circ}$ value of this angle is reached at the solar equator level (latitude 0$^{\circ}$).

The Polar Crown is formed in 1978 (see Fig. \ref{1670}), rises in latitude, and disappears at the pole in 1982 in Fig. \ref{1719}. This latitudinal migration is indeed the indication of cyclic variation as proposed by \citet[][Fig. 13]{Leroy-etal-84}. This conclusion is based on the available limited data time-span along with these 24 maps as presented here. Future work with a longer dataset may reveal this variation more comprehensively.

\section{Conclusion}
\label{conclusion}.

We publish these 24 synoptic maps of disambiguated averaged horizontal magnetic field vectors in 296 filaments or prominences in order to present in detail what has previously been synthetized in Fig. 5 of \citet{Leroy-etal-83} and Fig. 13 of \citet{Leroy-etal-84}, namely 1/ the field directions alternate from one neutral line to the next; 2/ the general field alignment is distorted from a solar north-south direction by the differential rotation effect, as was observed by \citet[][Fig. 2]{Hyder-65}, and is synthesized in Fig. \ref{azimuths} of the present paper. The small angle between the field vector and the long axis of the prominence appears to be systematic and could also result from the differential rotation effect, following also Fig. 13 of \citet{Leroy-etal-84}. The average value of this angle is 43$^{\circ}$ , as shown in Fig. \ref{histogram}. The differential rotation effect could in addition explain the north-south symmetry and oblique orientation of the long-axis directions of a prominence, as observed and depicted by \citet{Mazumder-etal-18} in their Fig. 6, even if the long-axis direction of the prominence coincides with the direction of the photospheric neutral line, which is not the direction of the distorted north-south magnetic field.

At the end of the observing period, the Pic-du-Midi polarimeter was able to quasi-simultenously observe the \ion{He}{i} D$_3$ and hydrogen H$\beta$ lines. Four polarization parameters were then measured, namely two polarization degrees and two linear polarization directions. Three of them correspond to the determination of the full magnetic field vector, whose horizontality was verified in 14 prominences. The remaining fourth parameter corresponds to the determination of an additional parameter, which was the electron density that causes the line depolarisation by collisions. The density was found to be on the order of $1 \times 10^{10}$ cm$^{-3}$, which is one order of magnitude weaker than generally thought based on analyses of the Stark effect. In addition, \citet{Stehle-etal-83} showed that the quasistatic approximation that is commonly applied to the Stark effect modeling is probably responsible for the overestimation of the electron density. They showed that results in better agreement with ours are obtained when ion dynamics is accounted when the Stark effect is modeled.

Upon concluding, we would like to emphasize how important it is to identify the observed prominence with a filament that is observed on the disk some days before or after, in order to properly evaluate the average scattering angle of the photospheric radiation by the prominence. Quiescent prominences are sufficiently high to enable departures of this angle up to $20^{\circ}$ from $90^{\circ}$, that is, from the so-called plane of the sky. We recall that the linear polarization degree behaves like $\mathrm{sin}^2\theta$, where $\theta$ is this angle \citep[][Eq. (37)]{SahalB-74}. Such a $20^{\circ}$ departure is then responsible for a more than 10\% of the relative change in the linear polarization degree, which is of great importance for the accuracy of a determination of the magnetic field strength and direction owing to the nonlinearity of the Hanle effect. This proper determination of the scattering angle from identification of the prominence with a filament is particularly important for Polar Crown prominences, which may even be observed above the solar pole. Such identification was not performed by \citet{Merenda-etal-06}, who observed a Polar Crown prominence and concluded that its average magnetic field was not horizontal, in contrast to the general result recalled in the present paper. Their observed polarization was incompatible with the horizontal field Hanle diagram plotted for right-angle scattering, but the scattering angle was not properly evaluated from identification with a filament. The observed polarization was not so far from the horizontal field Hanle diagram, however, and the box of measurement inaccuracies was not plotted either, which would also have led to the possibility of an average horizontal field in this prominence.

Another important point is resolving the field ambiguity. The magnetic field vector is not fully determined as long as the ambiguity is not resolved. The two ambiguous solutions may be as different as shown in Fig. 1 of \citet{Leroy-etal-84}. The lack of a resolved ambiguity is frequently masked in the present literature \citep{Schmieder-etal-13,Schmieder-etal-14a,Schmieder-etal-14b,Levens-etal-16a,Levens-etal-16b} by providing azimuth angle values within $180^{\circ}$ for the magnetic field vector. This vector is referred to in spherical coordinates with the $Oz$ -axis along the local solar radius. The azimuth is then defined from the projection of the magnetic field vector to the horizontal plane, and it is to be within $360^{\circ}$. When the scattering is right angle, the two ambiguous field vectors are symmetrical with respect to the line of sight, which is also the reference axis for the definition of the azimuth angle. The two ambiguous fields then have opposite azimuths. Therefore, only positive azimuths between $0^{\circ}$ and $180^{\circ}$ are provided in this literature without any ambiguity resolution. Moreover, when the scattering is not at right angle, the two ambiguous solutions are no more exactly symmetrical. Examples may be found in \citet{Bommier-etal-94}, where the average field strengths, inclinations, and azimuths of the two ambiguous solutions for 14 prominences are provided in Table III. It is shown there that the two ambiguous field strengths may differ by a factor up to two, under the effect of the departure from the right-angle scattering. This is again an example of the importance of properly evaluating the scattering angle from identification of the observed prominence with a filament observed on the disk some days after of before, as done in the present work. The proper evaluation of the scattering angle enables one of the methods used in the present work for solving the ambiguity, which is to compare the solution pairs obtained on two following days. The solar rotation effect modifies the symmetry of the ambiguous solutions.

\begin{acknowledgements}
We are grateful and indebted to an anonymous referee for an in-depth reading of our manuscript and for helpful suggestions increasing the interest of our work.
\end{acknowledgements}

\bibliographystyle{aa}

\end{document}